\documentclass{aa}

\usepackage{amsmath}
\usepackage{txfonts}
\usepackage{graphicx}
\usepackage{natbib}
\usepackage{amsfonts}
\usepackage{amssymb}
\usepackage{hyperref}
\usepackage{xcolor}
\usepackage{longtable}
\usepackage{appendix}

\usepackage{graphicx}
\usepackage{txfonts}

\newcommand{\numax}{\nu_{\mathrm{max}}}
\newcommand{\nucut}{\nu_{\mathrm{cut}}}
\newcommand{\teff}{T_{\rm eff}}
\newcommand{\yini}{Y_{\mathrm{ini}}}
\newcommand{\zini}{Z_{\mathrm{ini}}}
\newcommand{\fov}{f_{\mathrm{ov}}}
\newcommand{\tauheii}{\tau_{\mathrm{HeII}}}

\newcommand{\tdyn}{t_{\mathrm{dyn}}}
\newcommand{\dpone}{\Delta\Pi_1} 

\newcommand{\chifreq}{\chi^2_{\mathrm{freq}}}
\newcommand{\chispec}{\chi^2_{\mathrm{spec}}}
\newcommand{\chidyn}{\chi^2_{\mathrm{dyn}}}
\newcommand{\chicore}{\chi^2_{\Delta\Pi_1}}

\newcommand{\Kepler} {\textit{Kepler}\;}

\begin{document}

\title{Asteroseismology of 19 low-luminosity red giant stars from \Kepler}

\author{F. P\'erez Hern\'andez\inst{1,2}
        \and
        R. A. Garc\'{\i}a\inst{3}
        \and
        E. Corsaro\inst{1,2,3}
        \and
        S. A. Triana\inst{4}
        \and
        J. De Ridder\inst{5}
        }

\institute{Instituto de Astrof\'\i sica de Canarias, E-38205 La Laguna, Tenerife, Spain\\
       \email{fph@iac.es}
       \and
          Universidad de La Laguna, Dpto. Astrof\'\i sica, E-38206 La Laguna, Tenerife, Spain \\
       \and
          Laboratoire AIM, CEA/DRF-CNRS, Université Paris 7 Diderot, IRFU/SAp, Centre de Saclay, 91191 Gif-sur-Yvette, 
          France\\
       \and
          Royal Observatory of Belgium, Ringlaan 3, B-1180 Brussels, Belgium\\
       \and
          Institute of Astronomy, KU Leuven, Celestijnenlaan 200D, Leuven, Belgium\\
       }

\date{Received ; accepted April 28, 2016}

\abstract
{Frequencies of acoustic and mixed modes in red giant stars are now determined with high precision thanks to the long 
continuous observations provided by the NASA's \Kepler mission. 
Here we consider the eigenfrequencies of nineteen low-luminosity red giant 
stars selected by Corsaro et al. (2015) for a detailed peak-bagging analysis.}
{Our objective is to obtain stellar parameters by using individual mode frequencies and spectroscopic information.}
{We use a forward modelling technique based on a minimization procedure combining the frequencies of the p modes, the period
spacing of the dipolar modes, and the spectroscopic data.}
{Consistent results between the forward modelling technique and values derived from the seismic 
scaling relations are found but the
errors derived using the former technique are lower. The average error for $\log g$ is $0.002\,$dex, 
compared to $0.011\,$dex from the frequency of maximum power, $\numax$, and $0.10\,$dex from the spectroscopic analysis. 
Relative errors in the masses and radii are on average $2\%$ and $0.5\%$ respectively, compared to $3\%$ and $2\%$ derived 
from the scaling relations. No reliable determination of the initial helium abundances and the mixing length 
parameters could be made. Finally, for our grid of models with a given input physics, we found that low-mass stars require 
higher values of the overshooting parameter.}
{} 

\keywords{stars : red giants -- asteroseismology }

\maketitle

\section{Introduction}

The NASA's \Kepler mission \citep{2009IAUS..253..289B} and its recent version K2 
\citep{2014PASP..126..398H} are providing individual eigenfrequencies of a huge number of stars, including thousands of 
red giants (e.g. \cite{2013ApJ...765L..41S,2015ApJ...809L...3S,2015PASP..127.1038C}).
These data allow us to determine accurate stellar properties that help to constraint stellar evolution models 
\citep{2010ApJ...723.1583M,2012ApJ...749..152M} and
improve the determination of the properties of the exoplanets they might host 
\citep{2011A&A...527A..20G,2013ApJ...767..127H}.
A review of the progress with \Kepler in the field of the red giant stars can be found in
\cite{2012ASPC..462..503C} and \cite{GarStello2015}.

Obtaining stellar properties from pulsation spectra can be done with a variety of techniques such as forward modelling 
that seeks those models whose frequencies best match the observed ones, or inverse modelling which consists
in some (usually linearised) relation between the frequencies and the stellar structure.
Intermediate approaches take advantage of some analytical or asymptotic approximation to form
frequency combinations aimed at isolate some aspect of the stellar structure.
In this paper we use a forward modelling approach and compare the results with those obtained from simple scaling
relations and acoustic helium signature fits, the so-called acoustic glitches, e.g. 
\cite{1988IAUS..123..151V,1990LNP...367..283G}.
A similar comparison for main sequence and subgiants stars was done by \cite{2014ApJS..214...27M}. They found that 
the uncertainties in the masses and radii were improved by a factor 3 when individual frequencies were fitted 
compared to the use of empirical scaling relations.

The interest in making such a comparison is worth emphasizing:
on the one hand, the forward technique is model-dependent, 
introducing systematic errors into the stellar parameters but the
structural models and frequency computations can be done with an up-to-date physics.
On the other hand, the scaling relations have wide observational support 
(see \cite{2003IAUJD..12E..52B,2008ApJ...674L..53S}, and the specific work for read giants by
\cite{2013A&A...550A.126M}), but they are not fully understood from a theoretical point of view. 
This mainly concerns the relation between the frequency at maximum power, $\numax$, and the acoustical cut-off frequency, 
$\nucut$, which is supposed to justify the scaling relation $\numax \propto \nucut \propto  g /\sqrt{\teff}$, 
where $g$ is the surface gravity and $\teff$ the effective temperature.
Although some theoretical work has been done in an effort to understand this scaling relation 
\citep{2013ASPC..479...61B}, neither its extent or accuracy is clear.
In fact \cite{2015A&A...583A..74J} were able to measure the cut-off frequency for several stars and found a 
better agreement between $\numax $ and $\nucut$ than suggested by \cite{2013ASPC..479...61B}.
Thus both techniques should be regarded as complementary, and 
confronting their results enables us to gain confidence in the values and uncertainties of the stellar parameters derived.
The same is true when comparing results derived from forward modelling and a glitch fit, the latter being model-independent
but making use of an asymptotic approximation. 

The splitting of mixed modes show that red-giant cores rotate faster than their convective envelope
\citep{2012Natur.481...55B,2012ApJ...756...19D} and opens up
the possibility of probing their internal rotation rates \citep{2014A&A...564A..27D,2015A&A...580A..96D,
2016ApJ...817...65D}.
In fact, for our target stars \cite{2015A&A...579A..83C} determined the frequencies of many mixed dipolar modes including 
a high number of rotational splittings from which one can extract information on the internal rotation of this kind 
of low-luminosity red giant stars. We will analyse such information in a separate paper.

\section{Observations and Models}

\subsection{Observational data\label{sec_obser}}

\begin{table*}
\caption{List of target stars and their observed parameters. $\Delta\nu$ and $\numax$ are in $\mu$Hz. $\teff$, $\log g$, and 
$Z/X$ are obtained from spectroscopic observations. See text for details.}
\label{table_param}
\centering     
\begin{tabular}{r r c c c c c c}
\hline\hline  
KIC & & $\Delta\nu$ & $\numax$ & $\teff$ & $\log g$ & $Z/X$ \\
\hline             
 3744043 & A &  $ 9.90\pm 0.05$ & $112.5 \pm 0.2$ & $4906 \pm 91$ & $3.05\pm 0.11$ & $-0.37\pm 0.04$ \\
 6117517 & B &  $10.16\pm 0.05$ & $120.3 \pm 0.2$ & $4734 \pm 91$ & $3.01\pm 0.11$ & $ 0.38\pm 0.03$ \\ 
 6144777 & C &  $11.01\pm 0.06$ & $129.7 \pm 0.2$ & $4788 \pm 91$ & $3.07\pm 0.11$ & $ 0.24\pm 0.03$ \\ 
 7060732 & D &  $10.94\pm 0.05$ & $132.3 \pm 0.2$ & $4892 \pm 200$& $--          $ & $	--         $ \\ 
 7619745 & E &  $13.13\pm 0.07$ & $170.8 \pm 0.2$ & $4932 \pm 91$ & $3.13\pm 0.11$ & $-0.04\pm 0.03$ \\ 
 8366239 & F &  $13.70\pm 0.07$ & $185.6 \pm 0.4$ & $4948 \pm 91$ & $3.10\pm 0.11$ & $-0.00\pm 0.03$ \\ 
 8475025 & G &  $ 9.66\pm 0.05$ & $112.9 \pm 0.3$ & $4854 \pm 91$ & $3.01\pm 0.11$ & $-0.04\pm 0.03$ \\ 
 8718745 & H &  $11.40\pm 0.06$ & $129.3 \pm 0.2$ & $4769 \pm 91$ & $2.94\pm 0.11$ & $-0.32\pm 0.04$ \\ 
 9145955 & I &  $11.00\pm 0.06$ & $131.7 \pm 0.2$ & $4925 \pm 91$ & $3.04\pm 0.11$ & $-0.32\pm 0.03$ \\ 
 9267654 & J &  $10.34\pm 0.05$ & $118.6 \pm 0.2$ & $4824 \pm 91$ & $3.22\pm 0.11$ & $-0.04\pm 0.03$ \\ 
 9475697 & K &  $ 9.88\pm 0.05$ & $115.1 \pm 0.2$ & $4791 \pm 91$ & $2.90\pm 0.11$ & $ 0.19\pm 0.03$ \\ 
 9882316 & L &  $13.78\pm 0.07$ & $182.0 \pm 0.5$ & $5093 \pm 91$ & $3.20\pm 0.11$ & $-0.41\pm 0.04$ \\ 
10123207 & M &  $13.67\pm 0.07$ & $160.9 \pm 0.2$ & $4840 \pm 91$ & $2.98\pm 0.11$ & $-0.45\pm 0.04$ \\ 
10200377 & N &  $12.47\pm 0.06$ & $142.5 \pm 0.2$ & $4828 \pm 91$ & $3.00\pm 0.11$ & $-0.63\pm 0.04$ \\ 
10257278 & O &  $12.20\pm 0.06$ & $149.5 \pm 0.3$ & $4887 \pm 91$ & $2.99\pm 0.11$ & $ 0.06\pm 0.03$ \\ 
11353313 & P &  $10.76\pm 0.05$ & $126.5 \pm 0.2$ & $4955 \pm 91$ & $3.01\pm 0.11$ & $-0.42\pm 0.04$ \\ 
11913545 & Q &  $10.18\pm 0.05$ & $117.1 \pm 0.2$ & $4960 \pm 200$& $ --         $ & $	--	   $ \\ 
11968334 & R &  $11.41\pm 0.06$ & $141.4 \pm 0.3$ & $4826 \pm 91$ & $3.10\pm 0.11$ & $ 0.35\pm 0.03$ \\ 
12008916 & S &  $12.90\pm 0.06$ & $161.9 \pm 0.3$ & $5107 \pm 200$& $ --         $ & $  --         $ \\ 
 
\hline   
\end{tabular}
\end{table*}

The target stars considered in the present work and analysed by \cite{2015A&A...579A..83C} are 
listed in Table~\ref{table_param}. They are low-mass, low-luminosity red giant stars 
(specifically only stars with $\numax > 110\,\mu$Hz were considered) observed by \Kepler over more than 
four years and were selected because of their good SNR and 
the availability of gravity period spacing measurements from \cite{2012yCat..35400143M}.
Mode frequencies were obtained using the Bayesian tool \textsc{D\large{iamonds}} \citep{2014A&A...571A..71C}. 
Here one needs to implement a model
for the expected frequency pattern, and while the $\ell=1$ mixed modes were fitted individually, for each 
$\ell=2$ and $\ell=3$ peak only a single Lorentzian profile was used (see \cite{2012ApJ...757..190C} for more
details).
This means that the $\ell=2$ frequencies can be affected by the presence of mixed quadrupole modes and 
related rotationally split components. Unfortunately it is not possible to disentangle quadrupole mixed modes and 
rotational split components because of their high density in the $\ell=2$ frequency region.

We include frequencies from the nineteen stars analysed by \cite{2015A&A...579A..83C} but exclude
modes identified with a probability lower than $99\%$, as suggested by their peak significance test.
The large separations $\Delta\nu$ given in Table~\ref{table_param} are originally from \cite{2012yCat..35400143M} whereas the 
frequencies of maximum power, $\numax$, are a by-product of the peak bagging carried out by \cite{2015A&A...579A..83C}. 
On the other hand values of effective temperature ($\teff$), surface gravity ($\log g$), and surface metallicity ($Z/X$) were 
taken from the APOKASC Catalogue \citep{2014ApJS..215...19P} except for stars D, Q and S for which spectroscopic data were not 
available, and photometric values of $\teff$ from \cite{2012ApJS..199...30P} were used as input parameters. 

\subsection{Stellar code and pulsation}

Model fitting is based on a grid of stellar models evolved from the pre-main sequence with
the MESA code \citep{2011ApJS..192....3P},  version number 7184.
Models were computed with the OPAL opacities \citep{1996ApJ...464..943I}
and GS98 metallicity mixture \citep{1998SSRv...85..161G}.
Microscopic diffusion of elements was included; otherwise the standard MESA input physics was 
used. The choice of the metallicity mixture and its observed $Z/X$ value for the Sun was decided 
since as opposite to recent spectroscopic estimations 
(\cite{2005ASPC..336...25A,2009ARA&A..47..481A}), it gives a good agreement 
between solar standard models and helioseismic observations (see e.g. \cite{2008PhR...457..217B}).

The starting grid is composed of evolution sequences with masses ($M$) from $0.85M_{\odot}$ to $2.0M_{\odot}$,
initial helium abundances ($\yini$) from $0.25$ to $0.32$, initial 
metalicities ($\zini$) from $0.0025$ to $0.04$, and mixing length parameters ($\alpha$) from $1.5$ to $2.2$. 
The density of the original grid was increased to ensure that at least two values of every parameter were found in the 
solutions obtained after adding random noise to the data. This was checked for all the stars.
Specifically, the final mass step in the range $0.9\leq M\leq 1.5$ is $0.025M_{\odot}$ and  
the metallicity step for $Z\leq 0.025$ is $0.0025$. 
We do not found necessary to increase the original steps of, respectively, $0.01$ and $0.1$ for $\yini$ and 
$\alpha$.

No overshooting was considered in this global grid, but for each star, once other stellar parameters were fixed, 
new models were computed using the exponential prescription of \cite{2000A&A...360..952H}.
Here the particle spreading in the overshoot region can be described as a diffusion process with a diffusion coefficient
$D_{\mathrm{ov}}$ given by
\begin{equation}
D_{\mathrm{ov}} = D_0 \exp \left(\frac{-2z}{H_v} \right) \quad ; \quad H_{v}=f_{\mathrm{ov}} H_p
\; ,
\label{eq_over}
\end{equation}
where $z$ is the distance from the edge of the convection zone, $H_{v}$ the velocity scale height of
the overshooting convective elements at the edge of the convection zone, and $H_p$ the pressure scale height at the same point.
For the free parameter $\fov$, we have considered values from $0$ to $0.03$ in steps of $0.0006$.
The same $\fov$ value was used for the core and the envelope and throughout the evolutionary sequence.
Although, as specifically implemented in MESA, the formulation corresponds to that introduced by 
\cite{2000A&A...360..952H} to investigate the overshooting on AGB stars, it is in fact a simplified version of the
formulation given by \cite{1996A&A...313..497F}, who analysed the envelope overshooting in 
solar-like stars, main sequence A type stars 
and white dwarfs. Recently, \cite{2015A&A...580A..27M} carried out a seismic analysis of core overshooting in
a main sequence B star, obtaining satisfactory results compared to other overshooting formulations.

Frequencies were computed with ADIPLS code \citep{2008Ap&SS.316..113C}. The code uses the adiabatic approximation and 
neglects the interaction between convection and oscillations. Mode degrees from
$\ell=0$ to $\ell=3$ were considered. 

For a typical evolutionary sequence in the initial grid, 
we save between 100 and 200 models, from the subgiant phase to the red giant phase with an upper radius of about 
$8R_{\sun}$. Owing to the very rapid change in the dynamical time scale of the models, such grids are too coarse in the
time steps. Nevertheless, as detailed in Sec.~\ref{sec_tests}, we have checked that 
interpolations between models provide estimations of the p-mode frequencies,
the period spacing of the $\ell=1$ modes, and the stellar parameters with errors much lower than the 
observational ones. We have not attempted to fit single mixed modes; hence this procedure is safe and 
consumes relatively less time.

\section{Fitting procedure}

We have considered a $\chi^2$ minimization method including simultaneously mode frequencies and spectroscopic data. 
Specifically we minimize the function
\begin{equation}
\chi^2=\frac{1}{4} \left( \chifreq+ \chidyn + \chicore + \chispec \right)
\; .
\label{eq_chi2}
\end{equation}

Regarding the spectroscopic parameters, we have included when available the effective temperature ($\teff$) the surface 
gravity ($\log g$) and the surface metallicity ($Z/X$); namely:
\begin{equation}
\chispec= \frac{1}{3} \left[ \left( \frac{\delta \teff}{\sigma_{\teff}} \right)^2+ 
\left(\frac{ \delta (Z/X)}{\sigma_{ZX}} \right)^2+ 
\left(\frac{\delta g}{\sigma_{g}}\right)^2 \right]
\; ,
\end{equation}
where $\delta \teff$, $\delta (Z/X)$ and $\delta g$ correspond to differences between the observations and the models whereas
$\sigma_{\teff}$, $\sigma_{ZX}$ and $\sigma_{g}$ are their respective observational errors.

The other three terms in Eq.~\eqref{eq_chi2} are determined from the mode frequencies. The term $\chi^2_{\mathrm{dyn}}$ 
is aimed at minimizing the mean density through a term related to the large separation. In principle this term is not 
necessary in the minimization since the same information can be included in the term with the frequency differences. However, 
owing to the so-called surface effects not considered in the model and frequency computations, there will be some discrepancies 
between the large separation $\Delta\nu$ of the models and that of the observations. For instance, for the Sun, using a 
frequency interval around $\numax$ representative of our set of stars, we obtain for the mean difference between adjacent 
radial modes a value of $\Delta \nu = 134.8\,\mu$Hz for the observations and $\Delta\nu= 136.0\,\mu$Hz for a solar model; that 
is a difference of $0.9\%$. Although ultimately a solar calibration could be performed, hopefully cancelling some of the 
uncertainties, it seems better to fix the relevant constant in such a way that the discrepancy in the solar case 
is removed as far as possible.

Introducing the dynamical time, $\tdyn=(R^3/GM)^{1/2}$, and the dimensionless frequencies $\sigma_{nl}$ given by
$\omega_{nl}=\tdyn\,\sigma_{nl}$, the relative frequency differences between models and observations for radial modes 
can be expressed as
\begin{equation}
\frac{\delta \omega_{n0}}{\omega_{n0}} = \frac{\delta \tdyn} {\tdyn} + 
\frac{\delta \sigma_{n0}}{\sigma_{n0}} = \frac{\delta \tdyn} {\tdyn} + F(\sigma)
\; .
\label{eq_dimenfreq}
\end{equation}
Hence, one might expect that fitting the frequency differences for radial oscillations to a function of frequency, namely,
\begin{equation}
\frac{\delta \omega_{n0}}{\omega_{n0}} \to S^A (\omega) = A_0 + \sum_{i=1}^k A_i P_i(x)
\; ,
\label{eq_surfa}
\end{equation}
where $A_0$ is a constant, $P_i$ a Legendre polynomial of order $i$, and $x$ corresponds to $1/\omega$ linearly scaled to the 
interval $[-1,1]$, 
the constant term $A_0$ will be close to zero for models with the correct mean density. 
In what follows a value of $k=2$, corresponding to a parabolic function, has been adopted. We have checked 
that when values of $k=3,4\,$ are considered, the results for $A_0$ are the same to within the errors.

In practice, even for the best model, unknown surface effects will introduce a term in $F(\sigma)$, 
eventually including a constant that should be translated into the form of an 
uncertainty in the determination of $A_0$.
In fact, when considering frequency differences between the Sun and a solar model, 
and limiting again the range of radial orders around $\numax$ to that typically observed in the stars considered here,
we obtain $A_0=0.001$. In this way,
the error introduced in the solar mean density is almost an order of magnitude smaller than 
the discrepancy derived from the large separation computed as a simple average of frequency differences. It is worth mentioning
that, had we used the full range of known radial modes for the Sun, the surface term could be isolated by taking into account 
that at very low frequencies such terms tend to zero. 

For the minimization procedure we define the quantity
\begin{equation}
\chidyn = \left( \frac{A_0- A_{00}}{\sigma_{A_0}} \right)^2
\label{eq_chi2a0}
\; ,
\end{equation}
where $A_{00}$ is an offset caused by the simplified physics used in the model and frequency computations. In principle one
can take the solar value, $A_{00}=0.001$, but we did not found this offset completely satisfactory for our target stars and we
discuss it further on.
On the other hand $\sigma_{A_0}$ is the error associated to $A_0$ and in principle could also be taken as 
the discrepancy $\sigma_{A_0}=0.001$ found for the Sun.
We note that this value is at least one order of magnitude higher than the formal error found in a 
typical fit to Eq.~\eqref{eq_surfa}: hence, considering this higher uncertainty is the main reason for dealing with the 
terms $\chidyn$ and $\chifreq$ separately. 
In other words, and as suggested by Eq.~\ref{eq_dimenfreq}, $\chifreq$ will be intended for a minimization of the 
dimensionless frequencies. Its computation is detailed in the next paragraph. 

The term $\chifreq$ corresponds to the frequency differences of the $\ell=0,2,3$ modes after removing a smooth function
of frequency. The surface term is computed using only radial oscillations in a similar way to Eq.~\eqref{eq_surfa} except
that, as suggested by our tests (see Sec. \ref{sec_tests} below), the frequency differences were scaled with the 
dimensionless energy $I_{nl}$ (defined as in \cite{2010aste.book.....A}),
\begin{equation}
I_{n0} \frac{\delta \omega_{n0}}{\omega_{n0}} \to  S^B (\omega) = B_0 + \sum_{i=1}^k B_i P_i(x)
\; .
\label{eq_surfb}
\end{equation}
Here we have also adopted a value of $k=2$. Tests for some stars show that higher values do not significantly 
reduce the minimum value of $\chi^2$ while using $k=1$ gives substantially higher values.

Then, in the minimization procedure we consider radial as well as $\ell=2,3$ modes. 
The corresponding function to be minimized is
\begin{equation}
\chifreq = \frac{1}{N-k-1} \sum_{j=1}^N \left( \frac{ \delta\omega_j/\omega_j - I_j^{-1} S^B (\omega_j) }
{\sigma_{{\omega}_j}} \right)^2 
\; ,
\label{eq_minfreq}
\end{equation}
where $j$ runs for all the modes with degrees $\ell=0,2,3$, $N$ is the number of modes considered in the fit, and 
$\sigma_{{\omega}_j}$ the relative error in the frequency $\omega_j$.
We note that the polynomial function subtracted in Eq.~\ref{eq_minfreq} includes the constant coefficient, since 
a similar term (but with a very different uncertainty) was already included in $\chidyn$. 

As noted in Sec. \ref{sec_obser}, the observed modes include only one $\ell=2$ and one $\ell=3$ peak in every $\Delta\nu$ 
interval. To mimic the observations, at least to a first approximation, for each observed peak with $\ell=2,3$ 
we have taken an average of all the eigenfrequencies with the same degree and within a frequency interval of $2\,\mu$Hz 
around the mode with the lowest dimensionless energy and the correct `asymptotic' radial order,
weighted by $\overline{I_0}/I_{nl}$
where $\overline{I_0}$ is an interpolation of the dimensionless energy of the $\ell=0$ modes to the frequencies of 
the nonradial modes.
A value of $I_j=1/\sum I_{nl}^{-1}$ is assigned to the observed mode in Eq.~\eqref{eq_minfreq}. In practice, for modes with
degree $\ell=3$ this is basically equivalent to searching for the corresponding `pure' p mode, but for modes with degree 
$\ell=2$, an average between two modes is often required. 

The term $\chicore$ corresponds to differences in the period spacing of the $\ell=1$ modes, $\dpone$. 
However for the sake of rapidity and robustness in the computations, we have used a simpler parameter,
based on the work by \cite{2014MNRAS.444.3622J}. The main simplification compared to that work is that we fix the values of 
the hypothetical $\ell=1$ pure p-modes by using the frequencies of $\ell=0$ and $2$ modes, only fitting the coupling parameter 
and the period spacing that are assumed to be the same in the whole frequency range. 
We are not interested in using accurate equations for 
obtaining precise values of the period spacings but rather in using the same simple fit for the observations and the models.

The detailed computation is as follows. First we compute the small separation, 
$\delta\nu_{02}$ between adjacent $\ell=0$ and $\ell=2$ modes as an average over all available pairs. 
We then estimate the frequencies of the hypothetical pure $\ell=1$ modes with the 
asymptotic equation as $\nu^{p}_{n1}=(\nu_{(n+1)0}+\nu_{n0})/2-\delta\nu_{02}/3$.
Afterwards a first estimate of the period spacing is obtained with a linear fit of the dipolar period spacings,
$\Pi_{j}= 1/\nu_{j} - 1/\nu_{j+1}$ to a second order polynomial function of the variable:
\begin{equation}
y_j= \overline{ \left(\frac{\nu_j}{\Delta\nu}\right)^2 
\left[ q_0^2 + \frac{2}{3} \left( \frac{\pi \Delta\nu_j}{\Delta\nu}\right)^2 \right]^{-3/2} }
\label{eq_period_fit}
\end{equation}
where the bar denotes average of two consecutive dipolar modes, $j$ and $j+1$, 
$q_0$ is a guessed initial value for the coupling parameter $q$, and $\Delta\nu_j=\nu_j-\nu^p_{n1}$ where 
$\nu^p_{n1}$ is the closest pure p-mode to $\nu_j$. The period spacing $\dpone$ corresponds to the zeroth-order 
polynomial.\footnote{To derive Eq.\eqref{eq_period_fit} start from Eq.~(31)
in \cite{2014MNRAS.444.3622J} and note that $\delta x$ can be expressed in our notation as $\pi\Delta\nu_j/\Delta\nu$.
Also $\Delta\Pi$ here is $\Pi_j$ and $\omega_g=2\pi^2/\Delta\Pi_1$.
Then if $\Delta\nu$ is assumed to be known and a guess $q_0$ for $q$ is taken, this equation can be 
written as $\Pi_j \simeq \Delta\Pi_1 (1 + C \, \Delta \Pi_1 \, y_j)^{-1}$, where $C$ is a known constant.}
This value is used as an initial guess for a non linear fit. Introducing the global
parameters $\omega_g=2\pi^2/\dpone$ and $\omega_p= 2\Delta\nu$, and for every dipolar mode with angular frequency 
$\omega_j=2\pi\nu_j$ the functions $f_{1j}=q^2+2/3 (\pi \Delta\nu_j /\Delta\nu)^2$,
$f_{2j}=\omega_j^2 q^2/(\omega_p f_{1j}^{3/2})$, and $f_{3j}=1+ f_{2j}/\omega_g$, 
the period spacings are fitted to
\begin{equation}
\Pi_j \to \frac{2\pi^2}{\omega_g^2 f_3} \left(\frac{f_2}{\omega_g f_3} -1 \right)
\label{eq_period}
\; ,
\end{equation}
where $\omega_g$ and $q$ are the coefficients to be determined.

The corresponding minimization function is given by
\begin{equation}
\chicore = \left( \frac{\dpone (\mathrm{model}) - \dpone(\mathrm{observations}) } {\sigma_{\dpone}} \right)^2
\; ,
\label{eq_chi2pi}
\end{equation}
where $\sigma_{\dpone}$ is the uncertainty in the period spacing.
The formal errors resulting from the fit are too small and, if they were to be adopted, the minimization procedure based on
Eq.~\eqref{eq_chi2} would overweight the period spacing compared to the p-mode frequencies or the spectroscopic 
parameters.
This could cause the undesirable effect that many global and envelope structural properties become determined mainly by 
models with the correct $\dpone$. To prevent such situation, we have used the following criterion. First, we consider a high 
value of the uncertainty, namely $\sigma_{\Delta \Pi_1} = 10 \sigma_{A_0}$, in order to ensure that the other terms in 
Eq.~\eqref{eq_chi2} will give the proper global stellar parameters (more precisely, they will fix the model input parameters 
$M$, $\yini$, $\zini$ and $\alpha$). Then, we use a grid of models 
with different values of the overshooting parameter $\fov$ while introducing in Eq.~\eqref{eq_chi2pi}
the formal uncertainties for the period spacing, but still impose the condition
$\sigma_{\Delta \Pi_1} \geq \sigma_{A_0}$. 
We have checked that this worked as expected in the sense that some 
properties of the best fitted models, such as the core size or the age, are changed in the second iteration
whereas others, as the depth of the second helium ionization zone, remain almost unchanged.

\begin{figure}[!htb]
\centering
\includegraphics[scale=0.5]{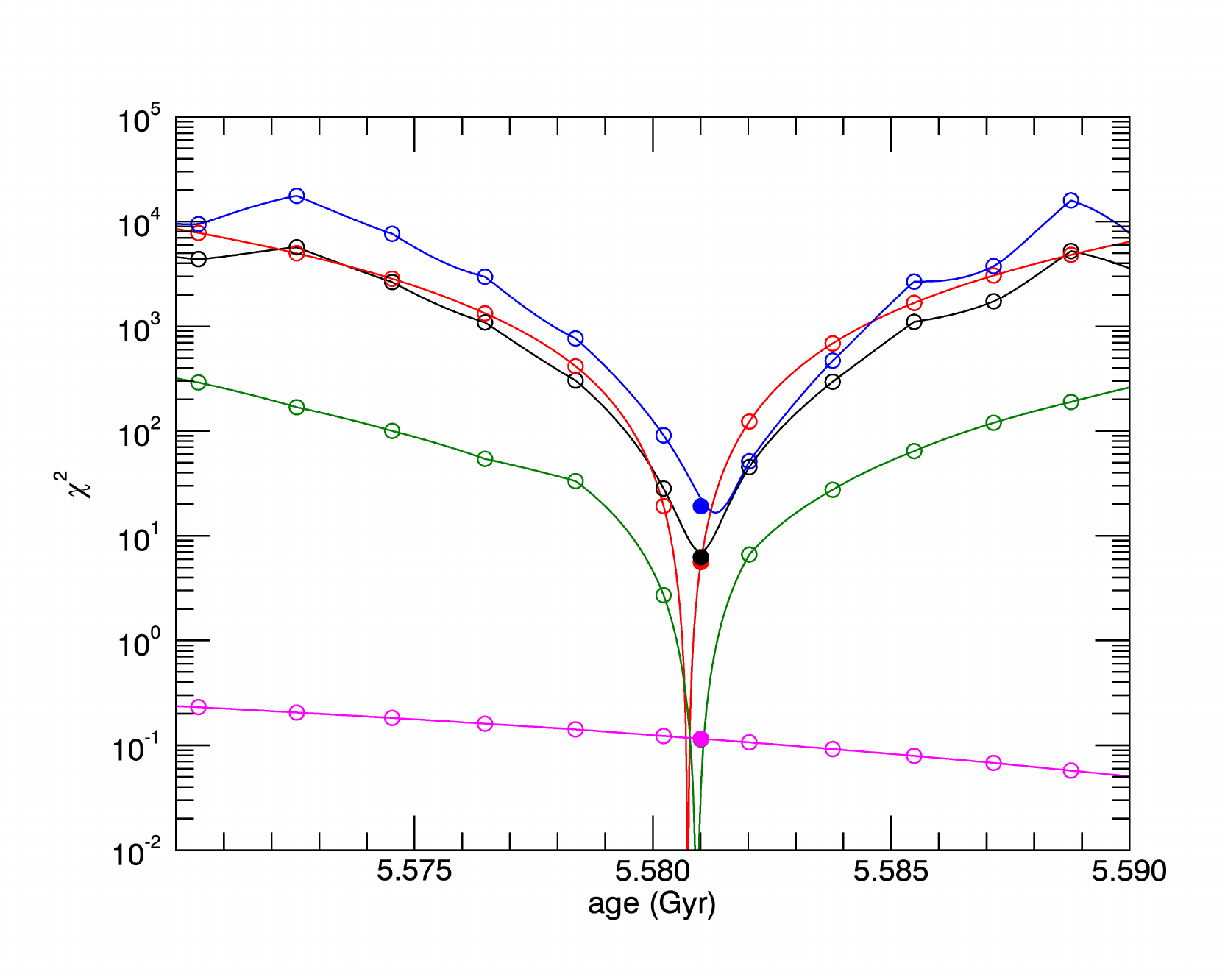}
\caption{Values of $\chi^2$ as function of age for an evolutionary sequence with stellar parameters that match 
those of KIC~003744043 within errors (letter A in tables\ref{table_param} and \ref{table_param2}).
The black open circles are values for models and frequencies actually computed whereas the continuous black line corresponds to 
parameters and frequencies interpolated to intermediate ages. The full black point is the $\chi^2$ value derived from 
the ``best fitted model'' explicit computed at the end of the procedure to exactly match the interpolated age. 
Also shown are values of $\chifreq$ (blue), $\chidyn$  (red), $\chicore$ (green), and $\chispec$ (magenta). Here the solid 
lines and the full and open points have the same meaning than before.}
\label{fig_interpol}
\end{figure}

\section{Tests on the methodology\label{sec_tests}}

As noted before, our minimization procedure is based on a grid that is too coarse 
for obtaining the best fitted models. Hence we interpolate their parameters and frequencies to a finer grid in ages.
We then search for the minimum $\chi^2$ corresponding to every evolutionary sequence.
We have found that the values of $\chi^2$ for our final models (which are 
explicitly computed from the pre-main sequence to the fitted age at the end of the minimization procedure) agree with the 
interpolated ones. As an example,
Fig.~\ref{fig_interpol} shows $\chi^2$ for an evolutionary sequence in the original grid with parameters 
matching those of KIC~003744043 (letter A in tables~\ref{table_param} and \ref{table_param2}). We show values for 
all the terms in Eq.\ref{eq_chi2}:
$\chi^2$ (black), $\chifreq$ (blue) $\chidyn$  (red) $\chicore$ (green), and $\chispec$ (magenta). 
The open circles correspond to models in the grid, the solid lines to values interpolated, and the full circles to the 
best fitted model. 
Since the example shown is for the best evolutionary sequence (within errors),  all the terms
have their minima at very close ages. To avoid misinterpretations, we note that this figure does not provide 
an indication of the uncertainty in the age since other parameter combinations can also give similar good fits but 
at different ages.

Although Fig.~\ref{fig_interpol} illustrates that the time step in our grid is fine enough, it can also be seen 
that the dependence of $\chifreq$ with age is not completely smooth, even in this very short range of ages. 
Including the individual frequencies of the mixed modes in the analysis with the goal of improving the results,
might result in more irregular $\chi^2$ functions than those shown in Fig.~\ref{fig_interpol} 
and hence further tests would be required to validate the procedure.

\begin{figure}[!htb]
\centering
\includegraphics[scale=0.5]{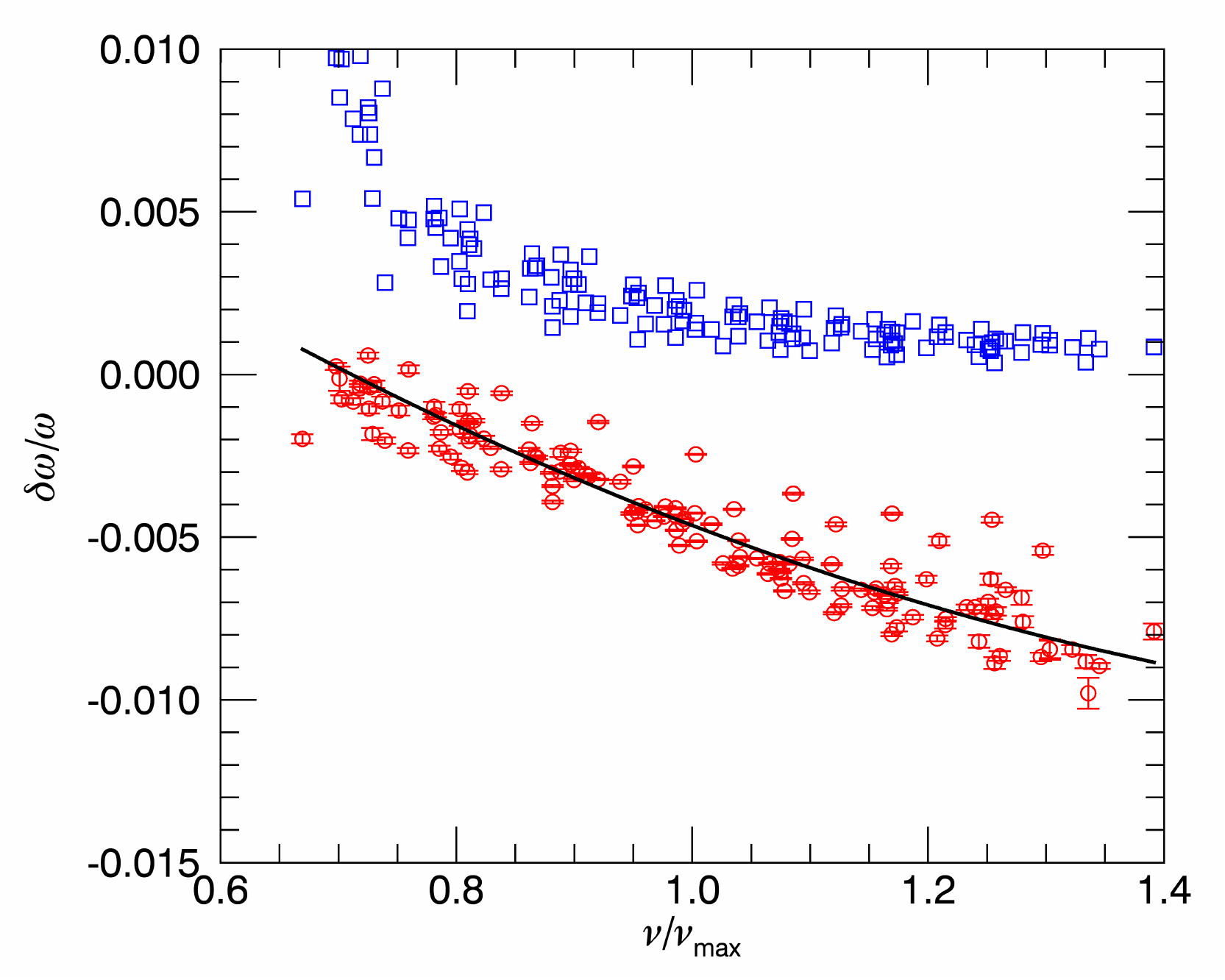}
\caption{Frequency differences between the observations and the best models, 
$[\omega ({\mathrm{observations}}) - \omega ({\mathrm{best~model}})]/\omega$ (red points), and associated 
dimensionless energies in arbitrary units (blue squares) for radial oscillations.
For a better comparison between modes in different stars, the horizontal axis gives the frequencies normalized to the 
frequency of maximum power. All nineteen stars are included. 
The black line is a second-order polynomial fit.}
\label{fig_dw_radial_all}
\end{figure}

Let us consider the offset $A_{00}$ introduced in Eq.~\eqref{eq_chi2a0}. As stated above, for the Sun we obtain $A_{00}=0.001$ 
but our tests show that using that number for the red giants gives rise to positive values of the frequency differences
$\omega ({\mathrm{observations}}) - \omega ({\mathrm{best~model}})$
in the low frequency range, at least for some of the stars. Since the 
missed surface effects are expected to overestimate the theoretical frequencies, it seems natural to impose the condition 
that the 
frequency differences will be negative 
for the whole frequency range, at least to within the dispersion of the minimization procedure.
For our sample of stars, a higher than solar value is required with a minimum offset of $A_{00}= 0.003\,$ to
$\,0.004$. We have finally taken $A_{00}=0.004$ and $\sigma_{A_0}=0.001$ in Eq.~\eqref{eq_chi2a0}. 
Alternatively, had we taken $A_{00}=0$ but explicitly impose the condition 
that the frequency differences cannot be positive, we would 
have found mean densities for our best models that differed from those reported here by a factor of $0.9986\pm 0.0015$. 
Since the source of error for the mean density is mainly that of $A_0$, this dispersion is consistent with the input 
value of $\sigma_{A_0}=0.001$ adopted as the estimate of the uncertainties in $A_0$.
We are aware that the constant shift depends on the physics of the models used (including the eigenfrequency computations), 
and also on the range of radial order spanned by the observed frequencies. Nevertheless, the former are the same and the 
latter very similar for all of our target stars. A comparison with a calibrated scaling relation will be given in 
Sec.~\ref{sec_results}.
In Fig.~\ref{fig_dw_radial_all} we show the frequency differences between our best models and the observations 
for all the radial modes and the full set of stars. 
As can be seen, when plotted against the normalized frequency $\nu/\numax$ they are similar for all the stars. 

\begin{figure}[!htb]
\centering
\includegraphics[scale=0.5]{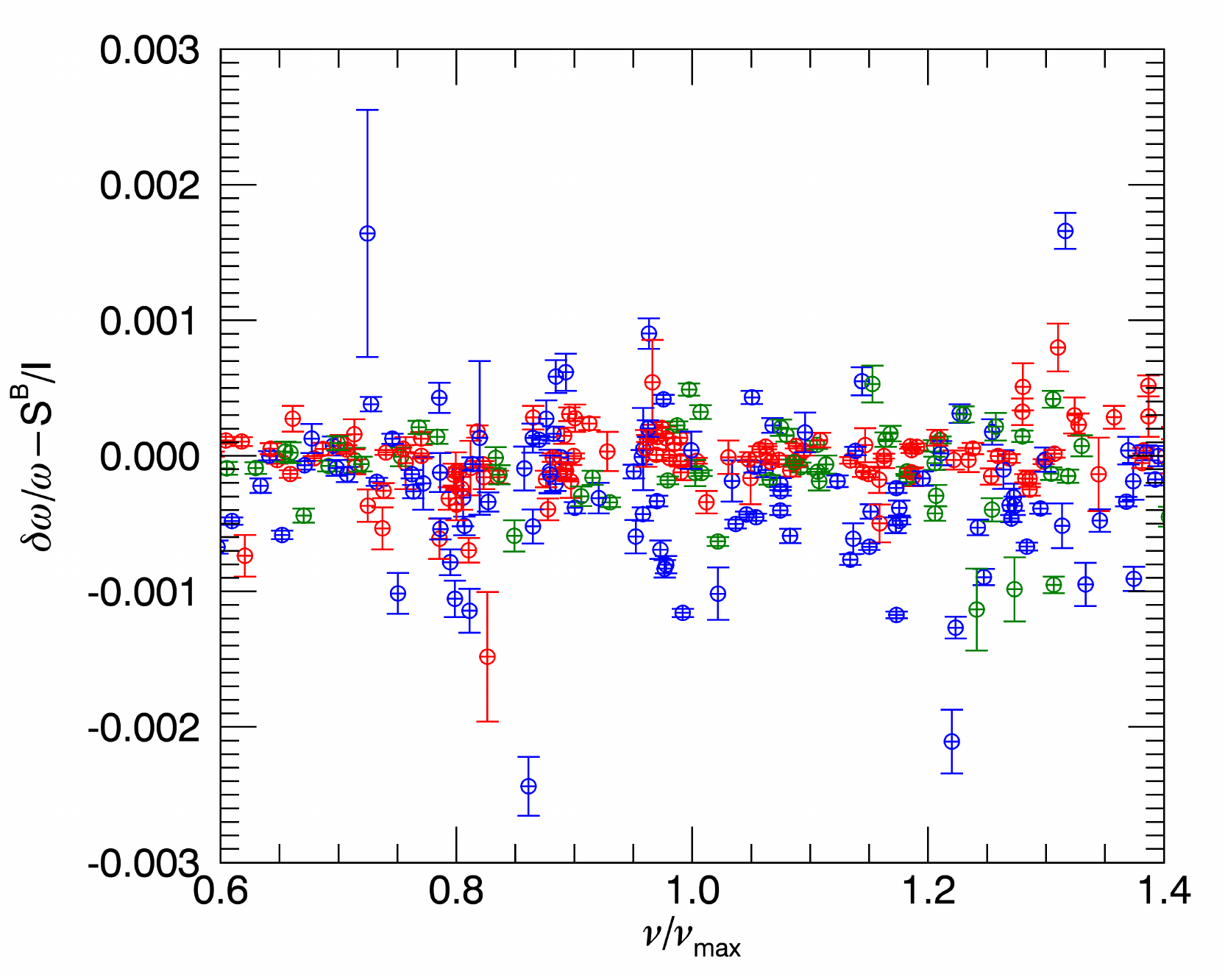}
\caption{Frequency differences between the observations and the best models after subtracting the surface effects. 
Red points are for $\ell=0$, blue points for $\ell=2$ and green points for $\ell=3$. All nineteen stars are included.}
\label{fig_dws_all}
\end{figure}

Figure~\ref{fig_dws_all} shows the residuals, that is, the relative frequency differences between the best models and the 
observations after subtracting the surface effects, for modes with degrees $\ell=0,2,3$ and all the stars.
The mean residuals for modes with degrees $\ell=0$ and $\ell=3$ is $0.0002$ whereas for modes with $\ell=2$ 
we obtain $0.0004\,$. These figures can be compared with the observational errors of $0.00008$,  $0.00010$, and $0.00007$
for modes with degrees $\ell=0, 2$ and 3 respectively. Thus for the $\ell=0$ and 3 our residuals are about two times higher 
than the observational errors whereas they are about four times higher for the $\ell=2$. The higher value for the $\ell=2$ 
modes is 
probably caused by the single peak considered in the fit to the observed spectrum to what is regularly a pair of 
quadrupolar modes of mixed character (prior to any rotational splitting consideration). 
The way we have dealt with this issue (see the paragraph after Eq.~\ref{eq_minfreq})
is therefore not enterly satisfactory.

Figure~\ref{fig_dw_radial_all} shows that the surface terms missing in the theoretical computations give rise to a smoother
frequency dependence in the relative frequency differences (red points) than that of 
the dimensionless energies (blue points) at the lowest frequencies,
indicating that the discrepancies between the theory and the observations should involve layers below the upper turning 
points of the lowest observed radial frequencies. It also suggests
computing $\chifreq$ by using a simple polynomial fit to 
$\delta\omega/\omega$ rather than to $I\delta\omega/\omega$. In that case the surface term will be computed using 
Eq.~\eqref{eq_surfa} while Eq.~\eqref{eq_minfreq} should be replaced by
\begin{equation}
\chifreq = \frac{1}{N-k-1} \sum_{j=1}^N \left( \frac{ \delta\omega_j/\omega_j - \overline{I_{0}}/I_j \, S^A (\omega_j) }
{\sigma_{\omega_j}} \right)^2 
\; ,
\end{equation}
where $\overline{I_0}$ is an interpolation of the dimensionless energy of the $\ell=0$ modes to the frequencies of 
the nonradial modes. If a second-order polynomial is considered, as in our reference case, we find residuals of 
$0.0003$ for the $\ell=$0 and 3 modes and $0.0006$ for the $\ell=2$, which is about $1.4$ times higher than when 
Eq.~\eqref{eq_minfreq} is used. We thus decided to keep using Eq.~\eqref{eq_surfb} for subtracting the surface effects.

\begin{figure}[!htb]
\centering
\includegraphics[scale=0.5]{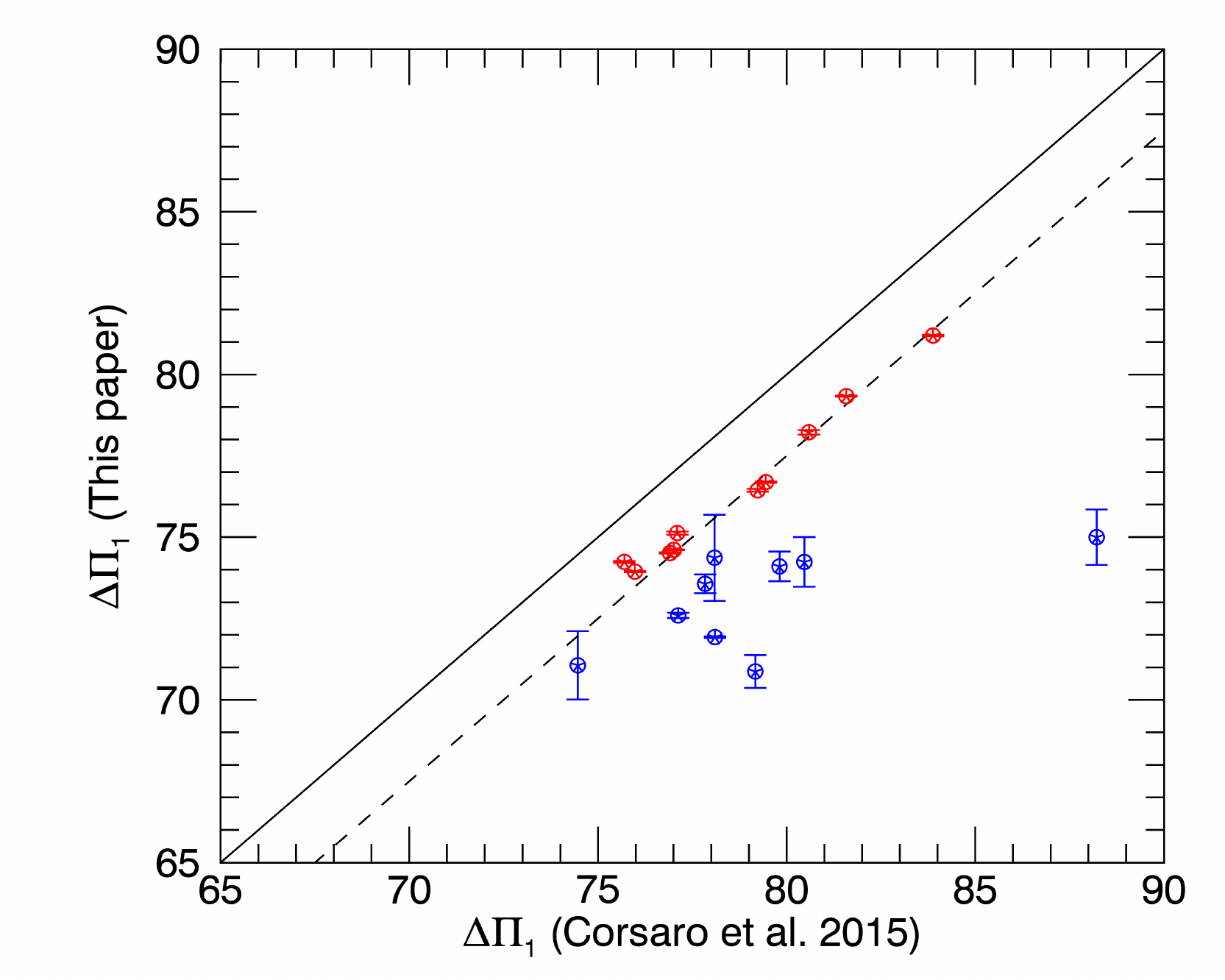}
\caption{Period spacing obtained in this work compared to that in \cite{2015A&A...579A..83C}.
Red points are values with errors lower than $0.1\%$. The continuous lines is $x=y$ and the dashed one $y=x-2.5\,$s.}
\label{fig_periods}
\end{figure}

As noted above,
the procedure used here for determining the period spacing is a simplified version of that in \cite{2014MNRAS.444.3622J},
and the resulting equation for the period spacing, Eq.~\eqref{eq_period}, is  
highly inaccurate compared to the observational errors.
Specifically we found that for the best models the individual periods deviate on average some $2.5\,$s from its fitted values.
Nevertheless we expect a similar behaviour for the observed modes and, hence, most of the differences will be cancelled out,
provided the observed and theoretical $\dpone$ values are computed in the same way.
In fact the average value for the standard deviations between the individual periods and their fitted values
is $4\,$s for the observations, $1.5$ times higher than the one derived from the best fitted models.
Figure~\ref{fig_periods} shows the period spacing for the nineteen stars computed with Eq.~\eqref{eq_period} against the 
$\dpone$ values given by \cite{2015A&A...579A..83C} (based on the works by \cite{2012yCat..35400143M} and
\cite{2014A&A...572L...5M}). 
Red points corresponds to those $\dpone$ values determined here with an error lower than $0.1\%$
The main difference is that our periods spacings are $2.5\,$s smaller. The blue point on the right with a period of 
$\Delta\Pi_1=88\,$s according to \cite{2015A&A...579A..83C} corresponds to KIC~008366239 for which we only used 6 modes in the 
fit. 

\section{Results\label{sec_results}}

\subsection{Individual stars}

\begin{table*}
\caption{Best models parameters. Age is given in Gyr and $\tauheii$ in seconds. The last line gives $1-\sigma$ absolute 
uncertainties except for $M$, $R$ and $L$ which are relative. The last two columns correspond to values obtained from
the scaling relations. $\yini$ and $\alpha$ are not well determined but their values are given for the
completeness in the model information.}
\label{table_param2}
\centering     
\begin{tabular}{r r r r r r r r r r r r | r r}
\hline\hline  
KIC        & & $M/M_{\odot}$& $\yini$ & $\zini$ & $\alpha$ & $\fov$ & Age   & $R/R_{\odot}$ & $L/L_{\odot}$ & $Z/X$ & 
$\tauheii$ & $M/M_{\odot}$& $R/R_{\odot}$ \\
\hline             
003744043 & A & 1.147 & 0.272 &  0.009  & 1.906 &  0.024 & 5.569 & 5.870  & 16.811 & 0.014 & 17710 & 1.16 & 5.88 \\
006117517 & B & 1.198 & 0.296 &  0.029  & 1.916 &  0.019 & 6.014 & 5.847  & 14.258 & 0.038 & 16780 & 1.21 & 5.86 \\
006144777 & C & 1.115 & 0.302 &  0.019  & 2.003 &  0.027 & 7.257 & 5.368  & 14.270 & 0.034 & 15630 & 1.11 & 5.41 \\
007060732 & D & 1.212 & 0.293 &  0.019  & 1.889 &  0.019 & 4.604 & 5.569  & 14.833 & 0.022 & 15330 & 1.25 & 5.65 \\
007619745 & E & 1.450 & 0.250 &  0.015  & 2.200 &  0.008 & 3.184 & 5.268  & 16.574 & 0.021 & 11790 & 1.32 & 5.09 \\
008366239 & F & 1.448 & 0.266 &  0.017  & 2.190 &  0.005 & 3.704 & 5.112  & 14.675 & 0.028 & 11180 & 1.43 & 5.08 \\
008475025 & G & 1.238 & 0.305 &  0.012  & 1.917 &  0.007 & 3.621 & 6.131  & 20.371 & 0.019 & 17590 & 1.27 & 6.16 \\
008718745 & H & 0.950 & 0.291 &  0.010  & 2.167 &  0.030 & 9.411 & 5.009  & 14.599 & 0.014 & 15870 & 0.96 & 5.02 \\
009145955 & I & 1.196 & 0.294 &  0.009  & 1.941 &  0.021 & 3.912 & 5.543  & 18.496 & 0.015 & 15170 & 1.22 & 5.58 \\
009267654 & J & 1.108 & 0.290 &  0.015  & 1.945 &  0.013 & 6.614 & 5.598  & 14.293 & 0.022 & 16910 & 1.11 & 5.63 \\
009475697 & K & 1.151 & 0.296 &  0.022  & 1.931 &  0.026 & 6.341 & 5.876  & 15.615 & 0.030 & 17660 & 1.20 & 5.96 \\
009882316 & L & 1.393 & 0.288 &  0.008  & 1.862 &  0.008 & 2.118 & 5.077  & 16.260 & 0.011 & 11090 & 1.38 & 5.00 \\
010123207 & M & 0.904 & 0.293 &  0.009  & 1.568 &  0.028 &11.666 & 4.373  &  7.931 & 0.014 & 12940 & 0.91 & 4.38 \\
010200377 & N & 0.943 & 0.273 &  0.005  & 1.757 &  0.028 & 8.302 & 4.703  & 12.334 & 0.007 & 14050 & 0.91 & 4.66 \\
010257278 & O & 1.249 & 0.250 &  0.020  & 2.199 &  0.006 & 6.337 & 5.254  & 14.460 & 0.028 & 13600 & 1.17 & 5.13 \\
011353313 & P & 1.250 & 0.260 &  0.005  & 2.200 &  0.017 & 3.375 & 5.716  & 22.539 & 0.007 & 15560 & 1.19 & 5.62 \\
011913545 & Q & 1.219 & 0.264 &  0.009  & 2.166 &  0.020 & 7.626 & 5.751  & 15.998 & 0.029 & 17250 & 1.19 & 5.82 \\
011968334 & R & 1.350 & 0.260 &  0.020  & 2.098 &  0.016 & 4.573 & 5.653  & 16.411 & 0.028 & 14380 & 1.27 & 5.52 \\
012008916 & S & 1.189 & 0.318 &  0.012  & 2.080 &  0.014 & 3.379 & 4.994  & 16.520 & 0.015 & 12420 & 1.27 & 5.08 \\
\hline             
$\sigma$  &   & 0.023 & 0.009 &  0.003  & 0.14  &  0.002 & 0.560 & 0.005  & 0.045  & 0.002 &    70 & 0.030 & 0.017 \\

\hline   
\end{tabular}
\end{table*}

Table~\ref{table_param2} summarizes the results for all the stars in our sample. 
To estimate the errors in the output parameters we 
add normally distributed errors to the observed frequencies, the coefficient $A_0$, and the spectroscopic parameters 
and search for the model with the minimum $\chi^2$ in every realization. 
In this way we estimate mean and $1\sigma$ values for $\chi^2$ and the stellar parameters. The last line in 
Table~\ref{table_param2} shows mean $\sigma$ values. 
Additionally we have redone all the analysis with a grid whose density along the mass axis is half the original
one. Compared to the full grid, a dispersion of $3\%$ in the mass is obtained as the mean for the nineteen stars.
Since other parameters are correlated with the masses, they also
change. In particular we obtain a mean dispersion of $1\%$ in the radii, $9\%$ in the luminosity, $0.96\,$Gyr in the age
and $177\,$s in $\tauheii$. These values are higher than those given in Table~\ref{table_param2} but they should 
be regarded as upper limits to the uncertainties introduced by the limited number of models in the grid.

Table~\ref{table_cova} gives the correlation matrix whose elements have been computed as averages 
of the linear Pearson correlation coefficients for all the stars in our dataset. Results are limited to the first 
step in the minimization procedure where only models without overshooting are considered and five parameters are
changed: $M$, $\yini$,  $\zini$, $\alpha\,$, and the age. 
The high correlation between $M$ and $Y$ has also 
been found in seismic analysis of main sequence stars (e.g. \cite{2014ApJS..214...27M}).
There is also a very high correlation between the age and $\zini$. 
Had we used the mean density rather than the age as the fifth parameter, the correlation coefficients on the last 
column would become very close to zero.

\begin{figure}[!htb]
\centering
\includegraphics[scale=0.5]{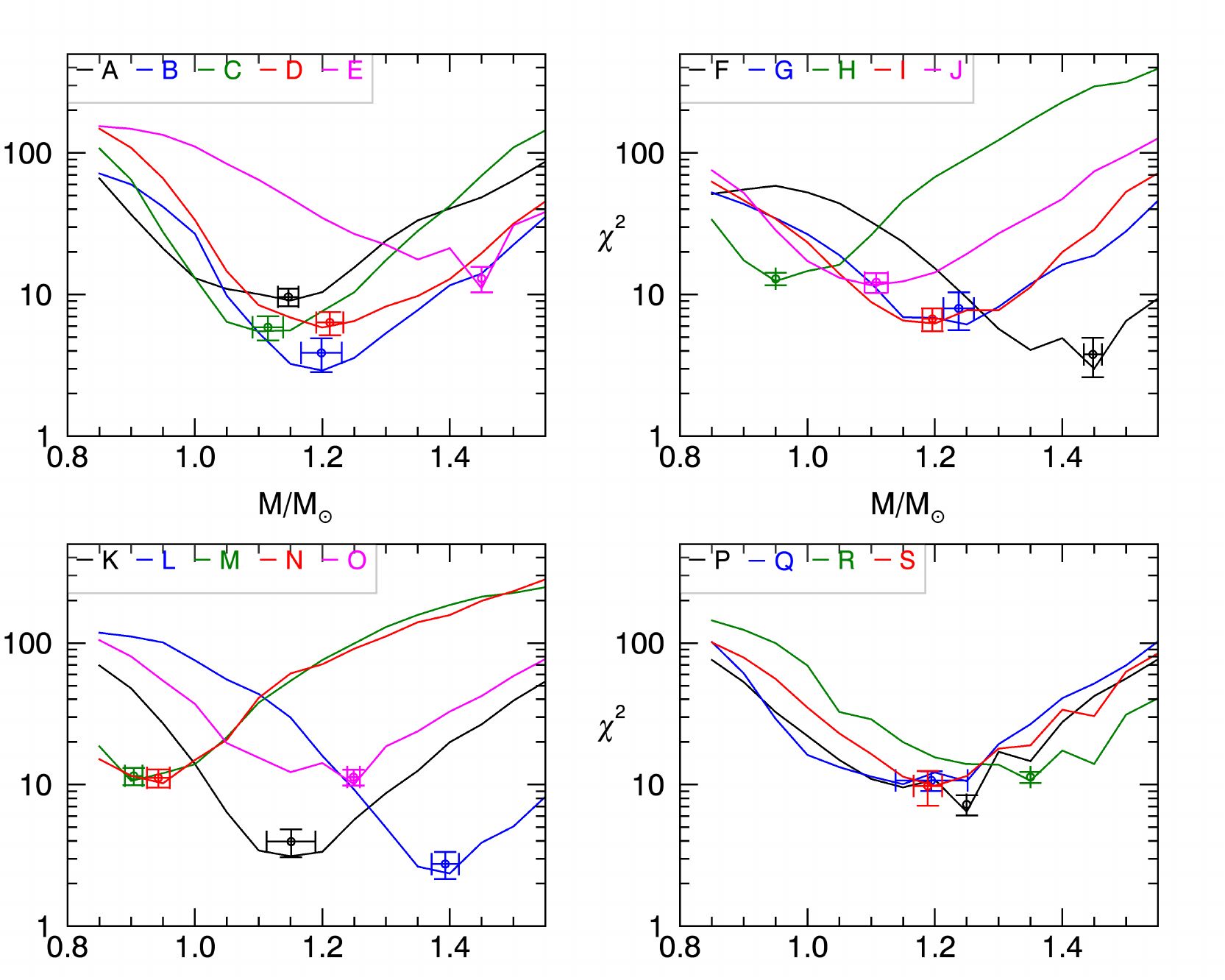}
\caption{Minimum values of $\chi^2$ as a function of $M$ --changing $\yini$, $\zini$ and $\alpha$--- for all the stars. 
Points with error bars correspond to $\chi^2$ values with $1\sigma$ uncertainties. 
The letters indicated in the legend correspond to those in Tables~\ref{table_param} and ~\ref{table_param2}.}
\label{fig_mass}
\end{figure}

\begin{figure}[!htb]
\centering
\includegraphics[scale=0.5]{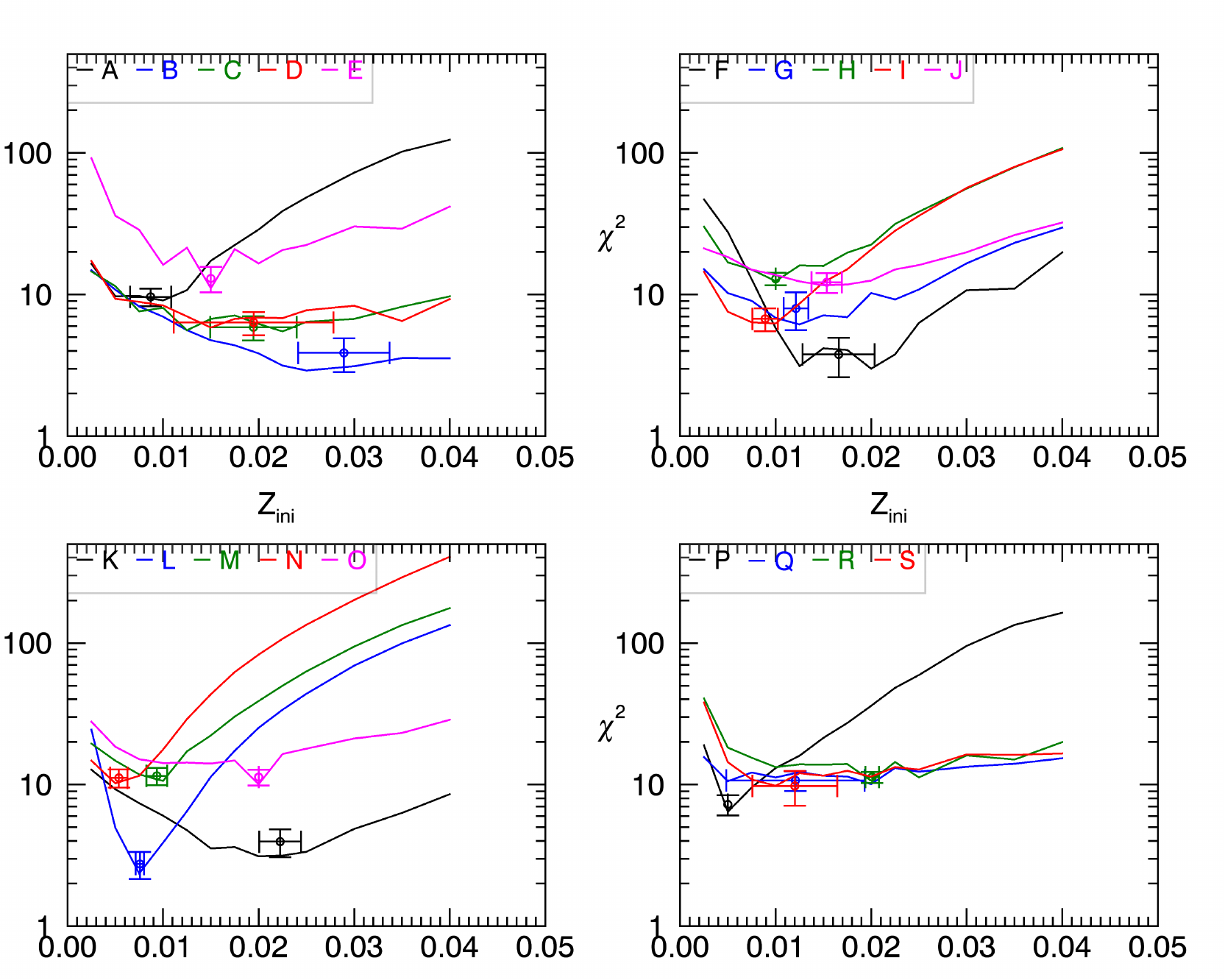}
\caption{Minimum values of $\chi^2$ corresponding to every $\zini$ for all the stars. 
Points with error bars correspond to $\chi^2$ values with $1\sigma$ uncertainties.
The letters indicated in the legend correspond to those in Tables \ref{table_param} and \ref{table_param2}.}
\label{fig_zini}
\end{figure}

\begin{figure}[!htb]
\centering
\includegraphics[scale=0.5]{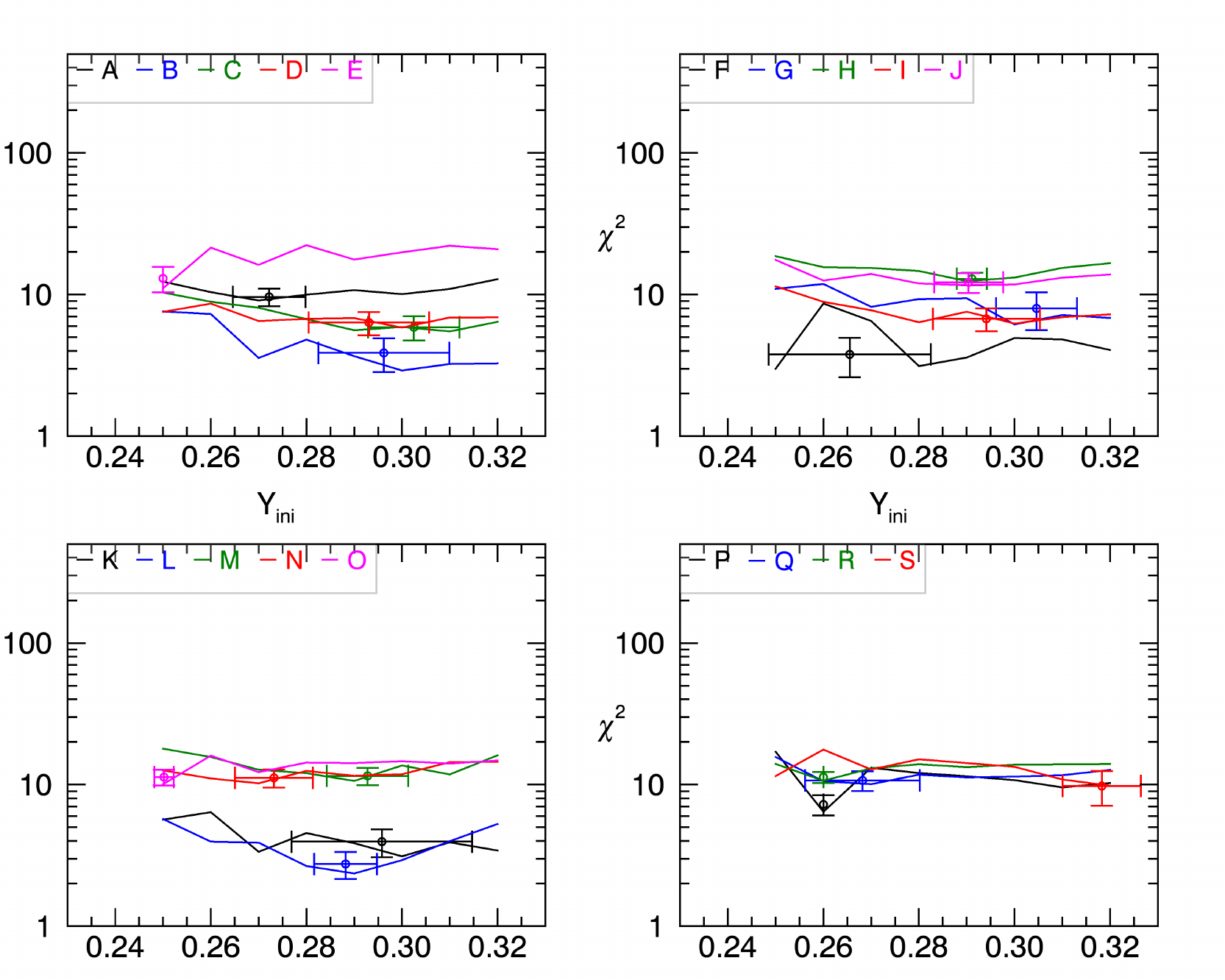}
\caption{Minimum values of $\chi^2$ corresponding to every $\yini$ for all the stars. 
Points with error bars correspond to $\chi^2$ values with $1\sigma$ uncertainties.
The letters indicated in the legend correspond to those in Tables \ref{table_param} and \ref{table_param2}.}
\label{fig_yini}
\end{figure}

\begin{figure}[!htb]
\centering
\includegraphics[scale=0.5]{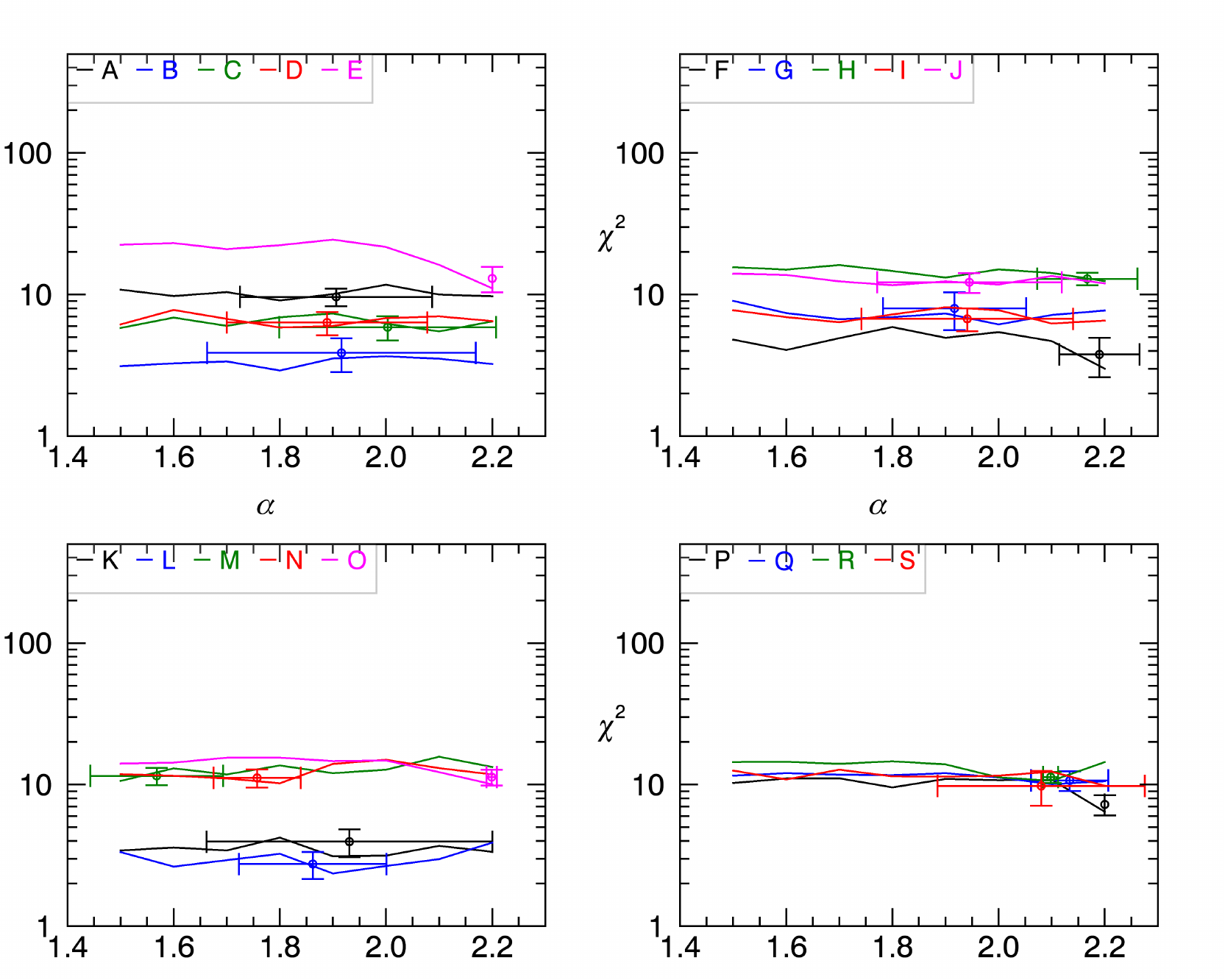}
\caption{Minimum values of $\chi^2$ corresponding to every $\alpha$ for all the stars. 
Points with error bars correspond to $\chi^2$ values with $1\sigma$ uncertainties.
The letters indicated in the legend correspond to those in Tables \ref{table_param} and \ref{table_param2}.}
\label{fig_alpha}
\end{figure}

Figure~\ref{fig_mass} shows $\chi^2$ as a function of $M$. For every mass we search for the minimum $\chi^2$
value by changing the remaining parameters ($Y_{ini}$, $Z_{ini}$ and $\alpha$). Models without overshooting were considered
here. The points with error bars correspond to the minimum $\chi^2$ value and $1\sigma$ uncertainty obtained as indicated
above. Figure~\ref{fig_zini} to \ref{fig_alpha} are similar but show $\zini$, $\yini$, and $\alpha$ 
respectively. In general the mass and the initial metallicity are well determined but the initial helium abundance 
and the mixing length parameter are not. 

\begin{table}[h]
\caption{Correlation matrix for the the grid of models without overshooting, corresponding to the 
first step in the minimization procedure.}
\label{table_cova}
\centering     
\begin{tabular}{c r r r r r}
\hline\hline  
        & $M$& $\yini$ & $\zini$ & $\alpha$ & Age\\
\hline 
$M$     &  1.00  & -0.71 & -0.25 &  0.13 & -0.39 \\
$\yini$ & -0.71  &  1.00 & -0.01 & -0.27 & -0.13 \\
$\zini$ & -0.25  & -0.01 &  1.00 &  0.22 &  0.78 \\
$\alpha$&  0.13  & -0.27 &  0.22 &  1.00 &  0.19  \\
Age     & -0.39  & -0.13 &  0.78 &  0.19 &  1.00  \\
 
\hline   
\label{table_cova}
\end{tabular}
\end{table}

\begin{figure}[!htb]
\centering
\includegraphics[scale=0.5]{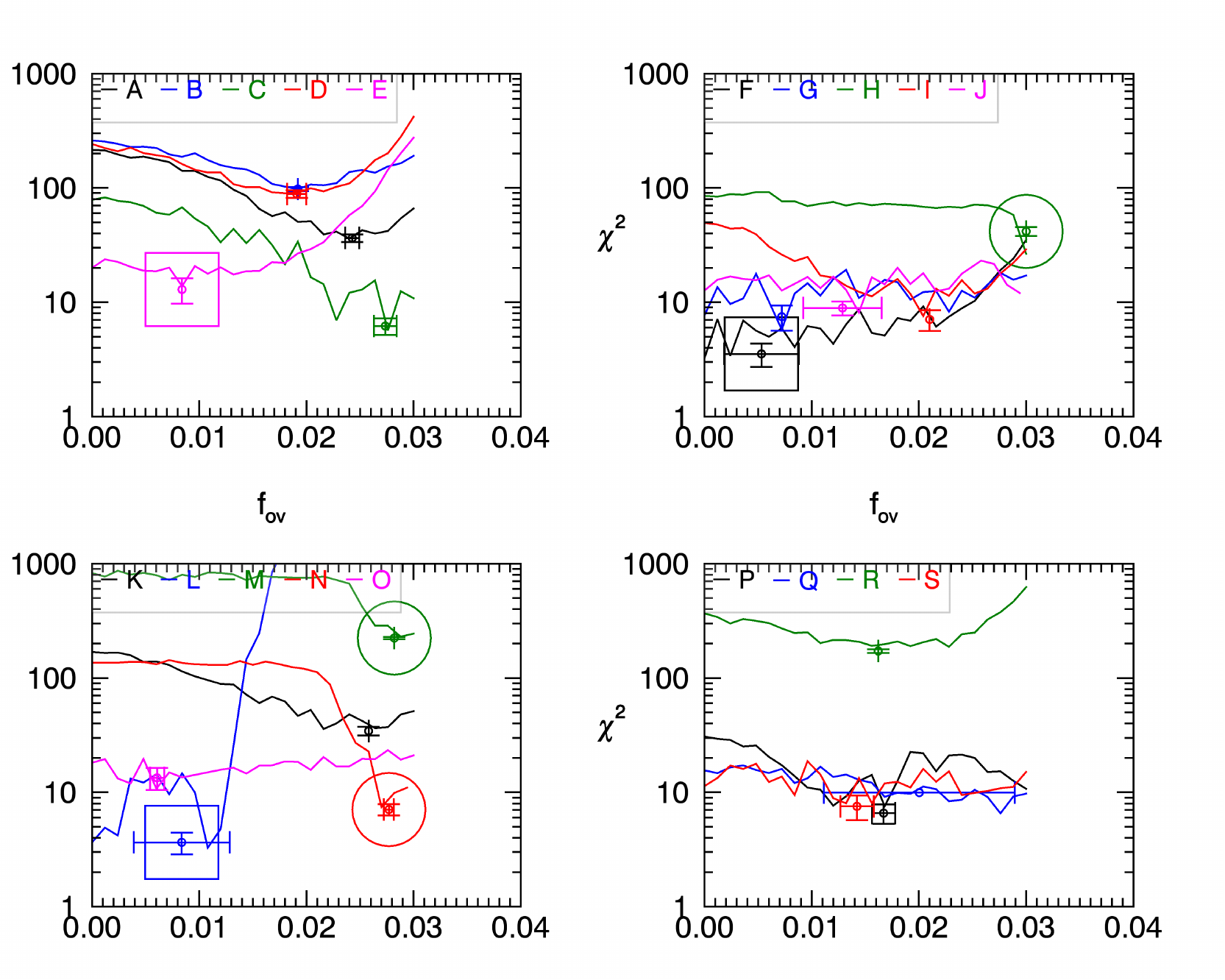}
\caption{Values of $\chi^2$ against the overshooting parameter $\fov$.
The points with error bars correspond to $\chi^2$ values with $1\sigma$ uncertainties.
The letters indicated in the legend correspond to those in Tables \ref{table_param} and \ref{table_param2}.
The big circles correspond to models with $M < 1M_{\odot}$ and the big squares are for models with $M>1.39M_{\odot}$.}
\label{fig_fov}
\end{figure}

As noted above, once the parameters $M, \zini, \yini, \alpha$ are determined by a $\chi^2$ minimization, we have considered
a grid of models with overshooting where $\dpone$ is properly included in the minimization function.
Figure~\ref{fig_fov} shows the corresponding $\chi^2$ values as a function of the overshooting parameter 
defined in Eq.~\eqref{eq_over}. Although $\fov$ is not always well determined we note that in some cases an upper or lower
limit can be inferred.

\subsection{Global Results}

\begin{figure}[!htb]
\centering
\includegraphics[scale=0.5]{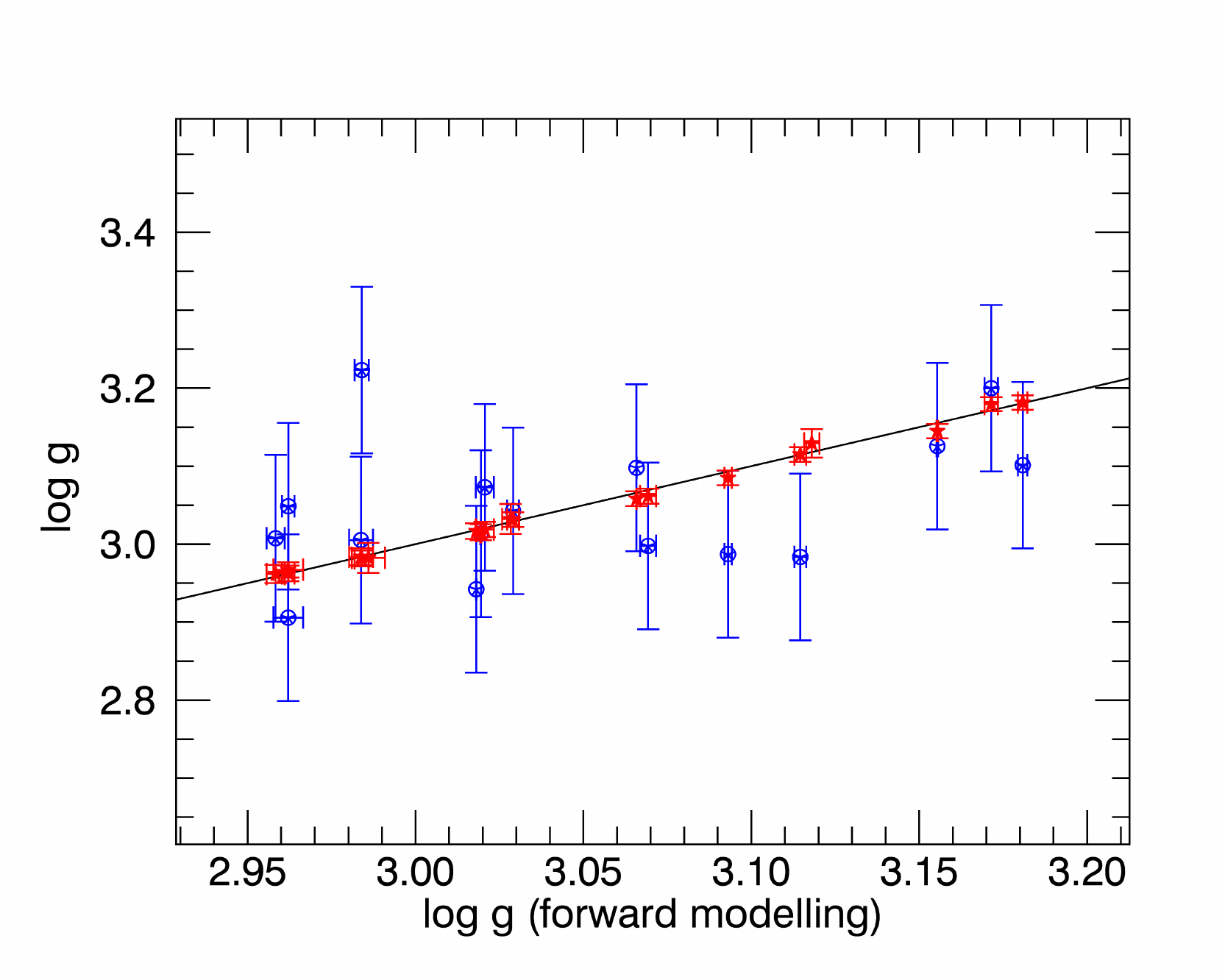}
\caption{Values of $\log g$ obtained in this work against the observed ones. The blue points correspond to the spectroscopic 
values and were used in the minimization procedure while the vertical coordinate of the red points were obtained by using the 
observed values of $\numax$ and $\teff$, assuming a relation $\numax \propto g/\sqrt{\teff}$. The observed values of 
$\numax$ were not used in the minimization procedure. The continuous black line corresponds to $x=y$.}
\label{fig_logg}
\end{figure}

As might be expected, whereas the input and output values for $\teff$ are basically the same, values of $\log g$ are 
improved once the asteroseismic information is included.
Specifically, the values of $\teff$ obtained in the minimization procedure are 
$75\pm 81\,$K higher than the input ones, which is consistent with the input and output errors of $108\,$K and $56\,$K 
respectively. 

On the other hand,
we obtain a mean error of $0.0018$ dex for the output values of $\log g$ whereas the input spectroscopic errors are 
$0.1$ dex on average.
In Fig.~\ref{fig_logg} we compare both values of $\log g$ for all the target stars. 
The results are consistent and we have not found any bias, the dispersion 
between both data being $0.09\,$ dex. Furthermore, in Fig.~\ref{fig_logg} we compare the values of 
$\log g$ obtained here with those derived from $\numax$ and 
$\teff$, assuming a relation of the form $\numax \propto g/\sqrt{\teff}$ and calibrated with the Sun. 
Values of $\numax$ and $\teff$ and their errors are taken from Table~\ref{table_param}. The resulting errors in
$\log g$ are on average $0.01\,$dex, about five times higher than that from the minimization procedure.
In any case, from Fig.~\ref{fig_logg} 
it seems clear that the agreement is much better in this instance, the mean difference being of 
$0.001\,$dex with a dispersion of $0.006\,$dex. 
We recall that we have not used the values of $\numax$ as input parameters, so the $\log g$ values obtained from
the minimization procedure and those from $\numax$ are observationally linked only through $\teff$.
Hence, this comparison proves the consistency between both methods and indicates that the formal output error of $0.002\,$ dex 
for $\log g$ obtained from the minimization procedure could be realistic.

The output values for the surface $Z/X$ found here was lower than the input spectroscopic values by 
$0.002 \pm 0.007$. This can be compared with the average input and output errors of $0.0035$ and $0.002$ respectively. 
Hence, we do not find any bias but the dispersion is about twice higher than expected.

\begin{figure}[!htb]
\centering
\includegraphics[scale=0.5]{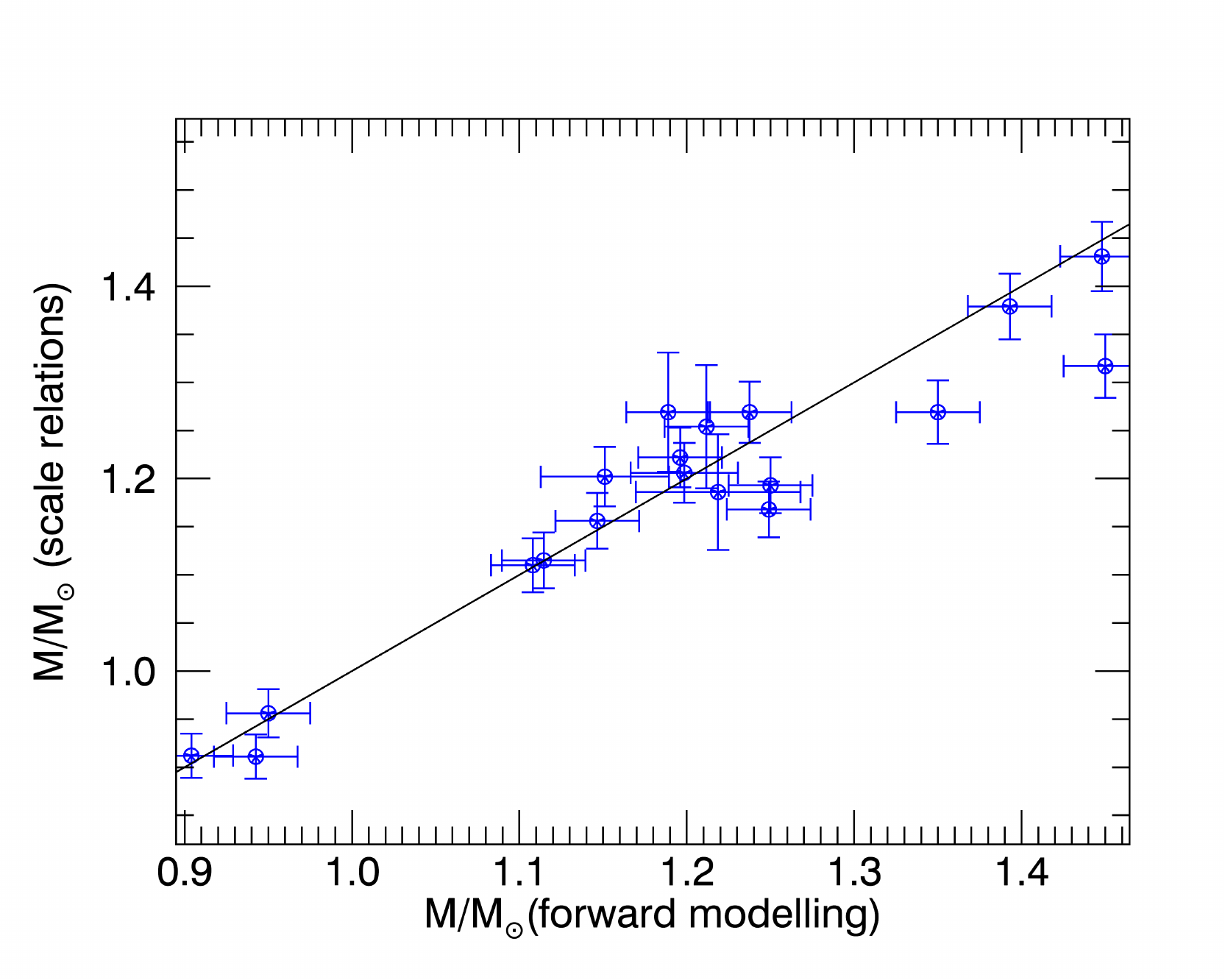}
\caption{Comparison between the masses obtained in the minimization procedure and those derived from simple scaling 
relations. The horizontal coordinate corresponds to the masses of the best models and the vertical coordinate gives the
asteroseismic values derived from the scaling relations.}
\label{fig_mass_mass}
\end{figure}

\begin{figure}[!htb]
\centering
\includegraphics[scale=0.5]{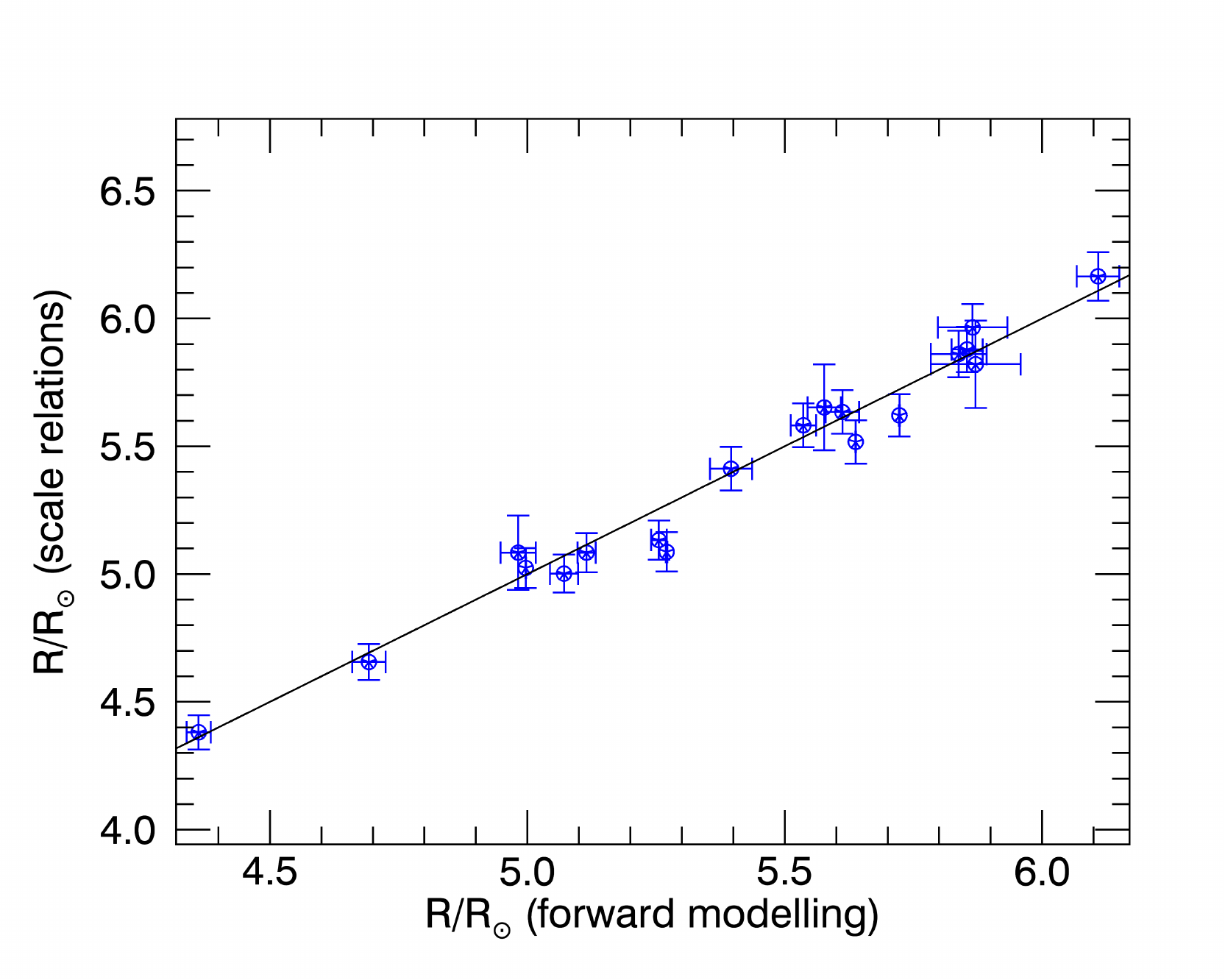}
\caption{Comparison between the radii obtained in the minimization procedure and those derived from simple scaling 
relations. The horizontal coordinate corresponds to the radii of the best models while the vertical coordinate gives the
asteroseismic values derived from the scaling relations.}
\label{fig_radii}
\end{figure}

The values of $M$ and $R$ determined here can be compared to those derived from the scaling relations.
Because for most of the stars we have used the spectroscopic values of $\teff$ from the 
APOKASC Cataloue \citep{2014ApJS..215...19P} whereas \cite{2012yCat..35400143M} used photometric determinations that are
on average lower by $145\,$K, the masses and radii derived here and those reported by \cite{2012yCat..35400143M} 
are also systematically shifted.
For a consistent comparison we have computed values of $M$ and $R$ from the scaling relations given by 
\cite{2013A&A...550A.126M} but using values of $\teff$, $\Delta\nu$ and $\numax$ from Table~\ref{table_param}.
Results for individual stars are given in the last two columns of Table~\ref{table_param2}, and
in Fig. \ref{fig_mass_mass} and \ref{fig_radii} we compare the values of $M$ and $R$ found here with those derived from
the scaling relations.
The masses derived from the minimization procedure are on average $0.006$ times lower than those derived from the scaling 
relations with a dispersion of $0.040$. On the other hand
the average relative errors are $0.023$ and $0.030$ for the 
the minimization procedure and the scaling relations respectively. 
Thus, both methods give consistent results, including their error determinations.
For the radii we found relative differences between the two methods of $0.002$ on average with a dispersion of $0.015$.
The formal relative errors are on average of $0.005$ and $0.017$ for the 
the minimization procedure and the scaling relations respectively. 
Hence, both methods again prove to be consistent.

\begin{figure}[!htb]
\centering
\includegraphics[scale=0.5]{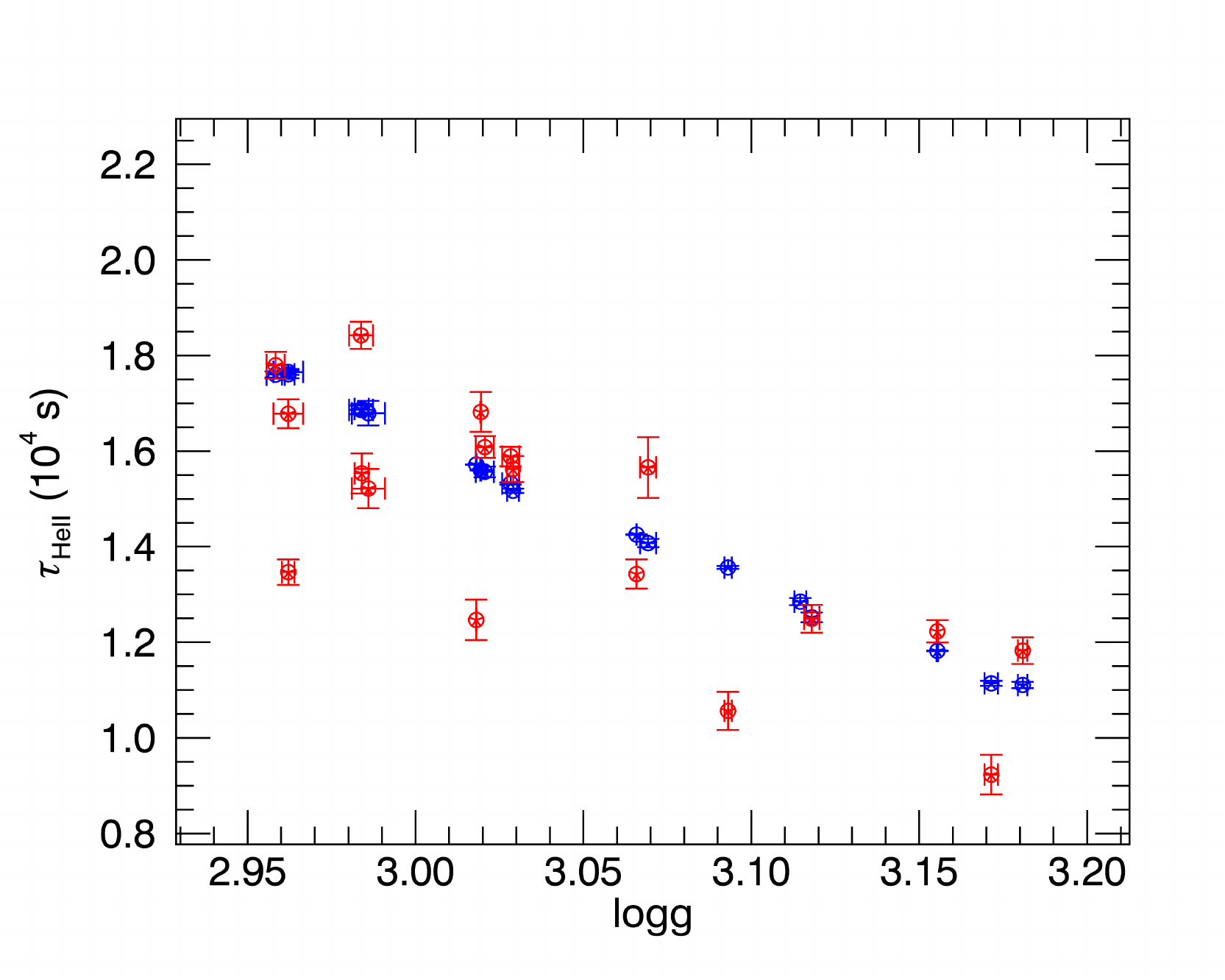}
\caption{Acoustic depth of the HeII ionization zone versus $\log g$. Blue points are for the best models and red points the
values derived by \cite{2015A&A...578A..76C} from a fit to the acoustic glitches.}
\label{fig_heii}
\end{figure}

It is also possible to compare the outputs from the forward modelling with those derived from the acoustic glitches.  
This is done in Fig.~\ref{fig_heii} where we show the acoustic depth of the second helium ionization zone, $\tauheii$, 
versus $\log g$. Blue points are for the forward modelling used in this work while the red points correspond to the values
derived by \cite{2015A&A...578A..76C} from a non linear fit of the second differences of the radial oscillations to a
model introduced by \cite{2007MNRAS.375..861H}.
As noted by \cite{2014MNRAS.440.1828B}, $\tauheii$ depends mainly on the dynamical state of the star and to a much lesser 
extend on the He abundance; hence, the simple relation between $\tauheii$ and $\log g$ shown in Fig.~\ref{fig_heii} for 
the results from the forward modelling. Values derived from the acoustic glitches are model-independent but
the fact that they do not reproduce such relation seems to indicate that errors in $\tauheii$ are higher than 
reported. Perhaps the problem arises because the number of measurements available is not much higher than 
the 5 free parameters of the theoretical equation considered. A test of different models and approaches to fit glitches 
could be considered in the future.

\begin{figure}[!htb]
\centering
\includegraphics[scale=0.5]{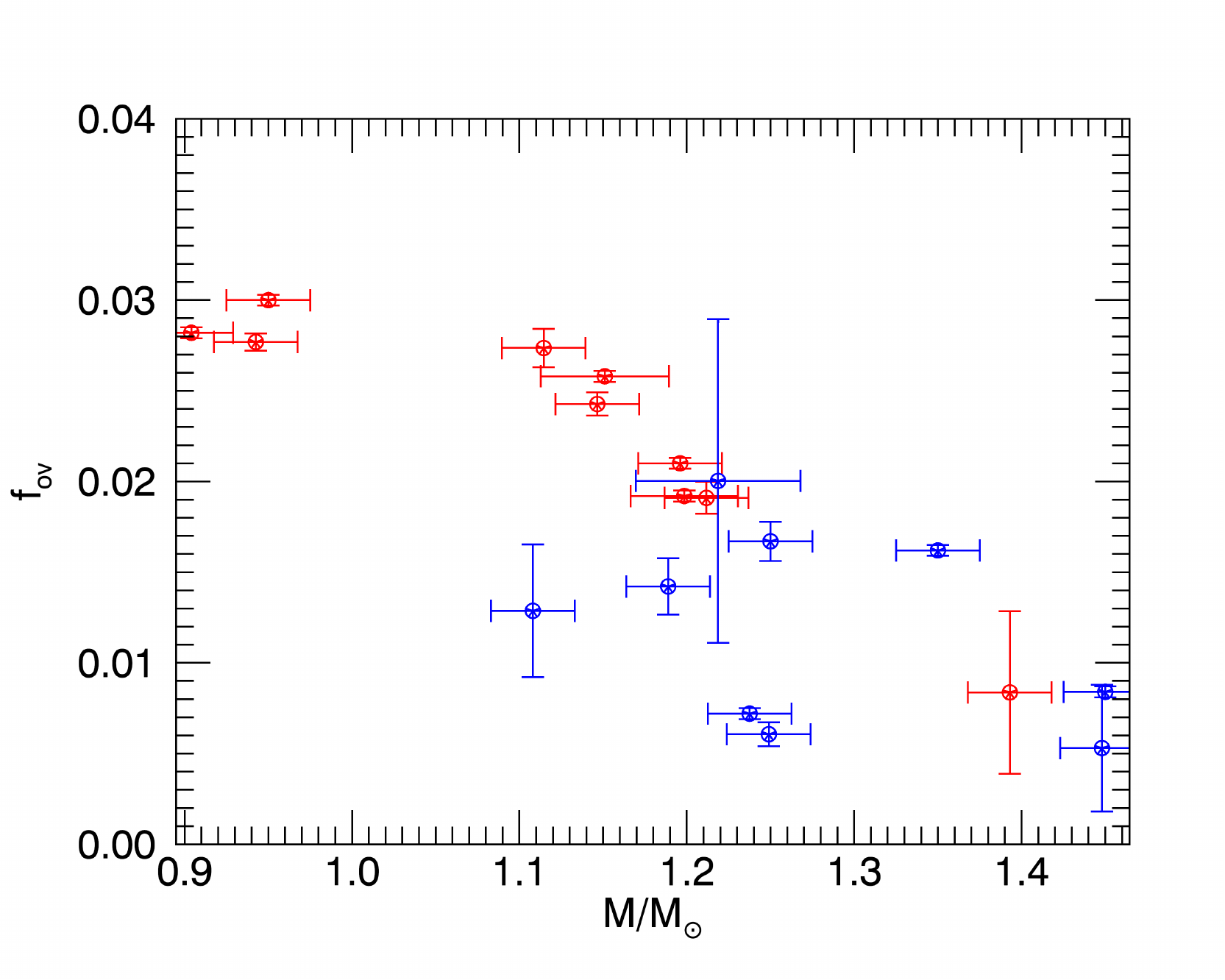}
\caption{Overshooting parameter versus mass. Red and blue points follows the same criterion as in Fig.~\ref{fig_periods}, the
red points corresponding to stars with a good determination of $\dpone$.}
\label{fig_over2}
\end{figure}

Finally in Fig.~\ref{fig_over2} we show the exponential overshooting parameter $\fov$ (see Eq.~\ref{eq_over}) versus the mass, 
$M$. Red points corresponds to the stars where $\dpone$ was obtained with a formal error lower than $0.1\%$ (the same 
red points 
than in Fig.~\ref{fig_periods}). It seems that there is some correlation between both parameters, lower masses corresponding 
to higher $\fov$ values. To gain and idea of the relevance of such relation we have identified in Fig.\ref{fig_fov}
the three stars with masses $M< 1M_{\odot}$ with big circles and the three stars with the highest masses with big squares. 
It seems clear that for all of our stars with $M< 1M_{\odot}$, values $\fov> 0.025$ give rise to a decrease in $\chi^2$  in
such a way that lower $\fov$ should be rejected for these stars. In a similar way, Fig.\ref{fig_fov} also indicates 
that for the more massive stars, the highest values of $\fov$ in the grid must be excluded. This seems more significant for
KIC~009882316 (letter L in Fig.~\ref{fig_fov}) 
which actually corresponds to the only massive star with a a good determination of $\dpone$ 
(the rightmost red point in Fig.~\ref{fig_over2}). Of course, this conclusion is model-dependent and the physical implications
are hard to extract.

\section{Conclusions}

We have used a forward modelling technique for obtaining stellar parameters of nineteen low-mass, low-luminosity red giant 
stars for which highly accurate frequencies are available thanks to the observations with \textit{Kepler}. 
In this first paper we have limited the work to the
p-mode frequencies and the period spacing of the $\ell=1$ modes. The relative frequency differences between our best models and 
the observations, once the surface effects are removed, are on average about $0.0002$, twice as high as the observational 
errors, for modes with degrees $\ell=0$ and 3  and about $0.0004$, four times higher than the observational errors, for the 
quadrupolar modes. The fact that these latter modes are worse fitted is probably caused by the regularly mixed nature of 
the $\ell=2$ which is hard to deal with properly, both observationally and theoretically.

The use of the p-mode eigenfrequencies and the spectroscopic values of $\teff$ and surface metallicities allowed to determine 
the masses and radii of the stars with uncertainties of $2\%$ and $0.5\%$ respectively. These figures can be compared 
with the $3\%$ and $2\%$ uncertainties derived from the scaling relations, that only use the global parameters 
$\Delta\nu$, $\numax$, and $\teff$. The consistency between both methods gives 
confidence in the individual values and the estimated errors of the stellar parameters reported in Table~\ref{table_param2}.
However, it should be noted that the forward modelling is not free of systematic errors due to the input physics and 
the methodology used. We have not attempted to estimate such uncertainties, but as a guide, for main-sequence and 
subgiant stars, \cite{2014ApJS..210....1C} estimated errors of $3.7\%$ and $1.3\%$ for the mass and radius respectively 
due to these factors. Given the agreement we have found between both methods, these figures seem rather high for our set of 
red giants. However this does not guarantee that other parameters as the age or $\tauheii$ would be affected
by the input physics and, hence, their errors can be higher than those given in Table~\ref{table_param2}.

On the other hand, other input parameters such as the initial helium abundance, $\yini$, and the mixing length parameter, 
$\alpha$, could not be unambiguously determined. 
In principle one might think that owing to the high accurate frequency measurements which allows us to detect clearly 
the presence of the glitch signatures caused by the second helium ionization zone \citep{2015A&A...578A..76C},
the helium abundance could implicitly be constrained by the forward frequency comparison between models and 
observations. However, one should take into account that isolating $\yini$ from other parameters is not a simple issue. 
In fact, in a related analysis for the solar case \cite{1994MNRAS.269..475P} found that changes in $\yini$ and the 
specific entropy of the adiabatic convection zone (parametrized by the mixing length parameter $\alpha$) are highly correlated.
It was only that for the Sun the known depth of the convection zone 
fixed the specific entropy that $\yini$ could be determined with low uncertainties by using the acoustic signatures
of the HeII zone.

We have found a correlation between overshooting and mass as shown in Fig~\ref{fig_over2}. However we are not claiming any
general physical implication for it. First, our grid of models with overshooting were limited to a second step in the 
search for the best models, and was only introduced once other parameters were fixed. We think this is reasonable for $M$ 
and $\zini$ but perhaps other parameters that were not well determined, such as $\yini$ and $\alpha$, should not be fixed in 
this second step. 
Second, the result is model-dependent and in particular only the prescription introduced by 
\cite{2000A&A...360..952H} was considered. Also, a change in the opacity tables or the metallicity mixture can give rise to 
different results. Such improvements, of course, would increase by some order of magnitudes the number of models to be 
considered.

\begin{acknowledgements}

This paper made use of the IAC Supercomputing facility HTCondor (http://research.cs.wisc.edu/htcondor/), partly financed by 
the Ministry of Economy and Competitiveness with FEDER funds, code IACA13-3E-2493.
E.C. is funded by the European Community's Seventh Framework Programme (FP7/2007-2013) under grant agreement 
n$^\circ$312844 (SPACEINN). R.A.G. acknowledges received funding from the CNES GOLF and PLATO grants at CEA and the ANR 
(Agence Nationale de la Recherche, France) program IDEE (n ANR-12-BS05-0008) ''Interaction Des 
\'Etoiles et des Exoplan\`{e}tes''.

\end{acknowledgements}

\bibliographystyle{aa}
\bibliography{biblio}

\end{document}